\documentclass[preprint,showpacs,preprintnumbers,amsmath,amssymb,aps,nofootinbib]{revtex4}

\usepackage{graphicx}% Include figure files
\usepackage{dcolumn}% Align table columns on decimal point
\usepackage{bm}% bold math
\usepackage{slashed}
\usepackage{hyperref}
\usepackage{color}
\preprint{JLAB-THY-13-1820}

\begin{document}

\title{Final-state interactions in inclusive deep-inelastic scattering \\
	from the deuteron}

\author{W. Cosyn}
\email{Wim.Cosyn@UGent.be}
\affiliation{Department of Physics and Astronomy,
	Ghent University, Proeftuinstraat 86, B-9000 Gent, Belgium}
\affiliation{Department of Physics, Florida International University,
	Miami, Florida 33199, USA}

\author{W. Melnitchouk}
\affiliation{\mbox{Jefferson Lab, 12000 Jefferson Avenue, Newport News,
	Virginia 23606, USA}}

\author{M. Sargsian}
\affiliation{Department of Physics, Florida International University,
	Miami, Florida 33199, USA}

\date{\today}

\begin{abstract}
We explore the role of final-state interactions (FSI) in inclusive
deep-inelastic scattering from the deuteron.  Relating the inclusive
cross section to the deuteron forward virtual Compton scattering
amplitude, a general formula for the FSI contribution is derived in
the generalized eikonal approximation, utilizing the diffractive
nature of the effective hadron--nucleon interaction.
The calculation uses a factorized model with a basis of three
resonances with mass $W<2$~GeV and a continuum contribution for
larger $W$ as the relevant set of effective hadron states entering
the final-state interaction amplitude.
The results show sizeable on-shell FSI contributions for Bjorken
$x \gtrsim 0.6$ and $Q^2 \lesssim 10$~GeV$^2$, increasing in magnitude
for lower $Q^2$, but vanishing in the high-$Q^2$ limit due to
phase space constraints.  The off-shell rescattering contributes at
$x \gtrsim 0.8$ and is taken as an uncertainty on the on-shell result.
\end{abstract}

% 11.80.Fv Eikonal approximation 
% 13.60.Hb Total and inclusive cross sections (including deep-inelastic
% processes)
% 24.85.+p 	Quarks, gluons, and QCD in nuclear reactions
% 25.30.-c 	Lepton-induced reactions

\maketitle

%%%%%%%%%%%%%%%%%%%%%%%%%%%%%%%%%%%%%%%%%%%%%%%%%%%%%%%%%%%%%%%%%%%%%%%%%
\section{Introduction} \label{sec:intro}

Inclusive deep-inelastic scattering (DIS) from the deuteron has for a
considerable time been the main source of information on the partonic
structure of the neutron \cite{Whitlow:1991uw, Benvenuti1990592,
Arneodo:1996kd, Arneodo:1996qe, Tvaskis:2010as}.
Recently there has been growing emphasis placed on such extractions
at large values of the momentum fraction $x$ carried by the partons
\cite{CJ10, CJ11, CJ12, Malace:2009dg, Martin:2012da, Ball:2013gsa}.
When combined with the more readily available proton data, and assuming
charge symmetry of the nucleon's parton distribution functions (PDFs),
one can directly reconstruct the individual $u$ and $d$ quark PDFs that
dominate nucleon structure at large values of $x$.

At high $x$ ($x \gtrsim 0.5$) the difference between the parton
structure of the proton and neutron grows and becomes rather
sensitive to the underlying QCD dynamics generating high-momentum
partons in the nucleon.
Unfortunately, in this region the extraction of parton distributions
in the neutron from inclusive deuteron DIS data becomes increasingly
complicated by the effects of nuclear corrections.
Within the nuclear impulse approximation, in which the scattering
takes place incoherently from individual nucleons bound in the nucleus,
these effects include nuclear Fermi motion and binding, relativistic
and off-shell corrections, and non-nucleonic components of the deuteron
wave function.
Considerable effort has been made over the years to understand these
effects quantitatively (see, {\it e.g.}, Refs.~\cite{Frankfurt:1976hb,
Kulagin:1994fz, Melnitchouk:1995fc, Afnan:2000uh, Afnan:2003vh,
Sargsian:2001gu, Kulagin:2004ie, Kahn:2008nq, Arrington:2011qt}).

Beyond the impulse approximation, rescattering effects can also play
an important role in the DIS process; for example, multiple scattering
of the beam from two (or more, for larger nuclei) nucleons can give rise
to nuclear shadowing corrections at small $x$ \cite{Frankfurt:1988zg,
Nikolaev:1990ja, Zoller:1991ph, Badelek:1991qa, Melnitchouk:1992eu,
Piller:1999wx}.
The interaction of the hadronic debris of the struck nucleon with the
spectator nucleon in the final state, on the other hand, can give
contributions also at higher $x$ values.
These final-state interaction (FSI) effects are generally considered
to be small in inclusive DIS at moderate to small values of $x$ ,
where, for large invariant mass $W$ produced in a high-energy collision,
the quantum phase space in the final state of the reaction is
unrestricted.
As a result, the closure relation can be applied to the sum over the
final states, enabling these to be represented through quark degrees
of freedom --- or, in other words, quark-hadron duality is expected
to hold in these kinematics \cite{Melnitchouk:2005zr}.
Dynamically, this is consistent with the picture in which the struck
quark hadronizes well after leaving the target, so that the final
state rearrangement does not influence the initial state probability
distribution of the interacting partons \cite{Collins:2011zzd}.

The situation can be quite different for large-$x$ and finite-$W$
kinematics, in which the condition for the closure approximation
(or duality) is not fully satisfied.  However, even if finite FSI
effects are expected here, estimating their contribution requires
knowledge about the composition and internal distribution of momentum
of the final hadronic state in the DIS process --- a problem which
remains very challenging.
The structure of the DIS final state can be examined by considering
the production of specific hadrons in coincidence with the scattered
electron \cite{Ciofi:2003pb, Ciofi:2000xj, Palli:2009it, Cosyn:2010ux,
Cosyn:2011jc}.  In particular, a promising avenue has been the study
of the distribution of slow tagged protons (with momentum up to
$\sim 500$~MeV/$c$) in semi-inclusive DIS (SIDIS) from the deuteron
\cite{Klimenko:2005zz, Baillie:2011za}.

Recently an approach was developed \cite{Cosyn:2010ux, Cosyn:2011jc}
for calculating the SIDIS reaction from the deuteron that accounted
for FSI effects based on general properties of high-energy diffractive
scattering.  The underlying assumption was that due to the restricted
phase space (finite values of $W$ and $Q^2$), the minimal Fock state
component of the wave function can be used to describe DIS from the
bound nucleon.  In this case the scattered state consists of three
outgoing valence quarks whose rescattering from the spectator nucleon
is parametrized in the form of a $Q^2$- and $W$-dependent diffractive
amplitude.  The results of this approach showed good agreement with
the most recent SIDIS data from Hall~B at Jefferson Lab
\cite{Klimenko:2005zz}, especially in the description of the rise
of the FSI effects in the forward direction of the spectator nucleon
production.  Comparison with the data also allowed one to extract
the parametric dependence of the diffractive rescattering amplitudes,
which increase with $W$ and decrease with $Q^2$.

A related calculation was performed in Refs.~\cite{Ciofi:2003pb,
Palli:2009it}, also using an eikonal approximation to estimate the FSI
amplitude.  Here a Glauber model was employed with a time-dependent
debris--nucleon cross section, including in addition the contribution
from the target fragmentation region to the SIDIS cross section,
and calculations for SIDIS off $^{12}$C were also presented.
Good agreement with the deuteron SIDIS data \cite{Klimenko:2005zz}
was found in a wide region of backward nucleon emission, whereas, as
expected, a traditional Glauber approach was difficult to accommodate
at forward spectator nucleon kinematics.  The kinematical region of
slow spectator protons in the backward hemisphere was found to have
small FSI contributions, making it useful for neutron structure
function extraction.  Fast protons in perpendicular kinematics,
on the other hand, yielded large FSI effects, making this region
suited for the study of hadronization mechanisms.  For fast spectator
protons the contribution from the target fragmentation region
was found to become significant, especially in the forward hemisphere.

Building on the knowledge gained from the semi-inclusive analyses,
in this paper we extend the approach of Refs.~\cite{Cosyn:2010ux,
Cosyn:2011jc} to inclusive DIS from the deuteron, over a similar
range of $Q^2$ and $W$ that was covered in the SIDIS kinematics.
The observation from the SIDIS studies \cite{Cosyn:2010ux,
Cosyn:2011jc} that the FSI structure is consistent with diffractive
scattering allows the generalized eikonal approximation (GEA) model
to be extended \cite{Cosyn:2010ux, Frankfurt:1996xx, Sargsian:2001ax}
to the inclusive DIS reaction through the optical theorem, relating
the inclusive cross section to the imaginary part of the forward
$\gamma^* D$ Compton scattering amplitude.
The general correspondence between the inclusive DIS cross section
and the forward Compton is derived in Sec.~\ref{sec:formalism}.
The Compton scattering amplitude is then computed, firstly in the
plane-wave Born approximation, and then in the presence of final-state
hadronic interactions, taking into account both on-shell and off-shell
contributions in the rescattering amplitude.
The results presented in Sec.~\ref{sec:formalism} are rather general,
relying only on the diffractive nature of the FSI.  The specific model
used to obtain the numerical estimates of FSI effects is introduced in
Sec.~\ref{sec:estimates}, where its main assumptions and approximations
are highlighted.  These include a factorized approach for the hadronic
currents in the FSI amplitude, and a three-resonance model combined 
with a DIS continuum region distribution at large $W$ for the states
that contribute to the FSI.  The numerical results for the FSI effects
are presented in Sec.~\ref{sec:results}, and conclusions are drawn
in Sec.~\ref{sec:conclusion}.

%%%%%%%%%%%%%%%%%%%%%%%%%%%%%%%%%%%%%%%%%%%%%%%%%%%%%%%%%%%%%%%%%%%%%%%%%
\section{Theoretical framework} \label{sec:formalism}

In this section we present the definitions for cross sections and the
nuclear hadronic tensor corresponding to the inclusive scattering of
an electron $e$ from a nucleus $A$,
\begin{equation} \label{eq:reaction}
 e(k_i) + A(p_A) \rightarrow e'(k_f) + X(p_X)\, ,
\end{equation}
where $k_i$ and $k_f$ are the four-momenta of the initial and final
state electrons, and $p_A$ and $p_X$ are the four-momenta of the
target nucleus and the produced hadronic system $X$, respectively.
While the formal results derived here will be valid for any nucleus
$A$, in the actual calculations we will specialize to the case of the
deuteron.

We base our derivations on the relationship between the inclusive 
electroproduction cross section and the imaginary part of the amplitude 
of forward virtual Compton scattering off the nucleus.  The advantage
of such an approach is that the amplitudes accounting for the FSI
effects will self-consistently satisfy the unitarity conditions for
inelastic rescattering.  An alternative approach would be to introduce
FSI effects in the $\gamma^* D \to X$ scattering amplitude and apply
AGK type cutting rules \cite{Abramovsky:1973fm} in the calculation of
the cross section to restore unitarity.
In our approach we explicitly identify the Born and FSI terms of
the inclusive electroproduction cross section with the impulse
approximation and FSI contribution in the forward Compton scattering
amplitude, with the latter calculated in the GEA.

% ........................................................................
\subsection{Inclusive cross section and forward nuclear virtual
	Compton scattering}
\label{subsec:gen}

Neglecting electron masses, we define the differential DIS
cross section as
\begin{equation}
 d\sigma
= \frac{1}{4\sqrt{(k_i\cdot p_A})^2}
  \sum_X \overline{\sum_{i,f}}\
  |\mathcal{M}|^2\,
  (2\pi)^4\, \delta^{(4)}(q+p_A-p_X)
  \frac{d^3{\bm k}_f}{(2\pi)^3\, 2\epsilon_f}
  \frac{d^3{\bm p}_X}{(2\pi)^3\, 2E_X},
\end{equation}
where $\epsilon_f$ and $E_X$ are the energies of the final electron and
hadronic state $X$, $q = p_X - p_A = k_i - k_f$ is the four-momentum
transfer to the target, and we average the square of the scattering
amplitude $\mathcal{M}$ over the initial spins of the electron and
nucleus and sum over the scattered electron spins.
The formal sum $\sum_X$ includes all possible final states $|X\rangle$
and integrates over the distributions of their internal momenta.
Using the phase space identity for $p_X$,
\begin{equation} \label{eq:delta}
 \frac{d^3{\bm p}_X}{2E_X}\
=\ d^4p_X\, \delta\left(p_X^2-W_X^2\right)\, \theta(E_X),
\end{equation}
we can eliminate the four-dimensional $\delta$-function with
$d^4p_X$, and express the differential cross section as
\begin{equation} \label{eq:crossA}
 \frac{d\sigma}{d\epsilon_fd\Omega_f}
= \frac{1}{(4\pi)^2} \frac{1}{2M_A} \frac{\epsilon_f}{\epsilon_i}\,
  \sum_X \overline{\sum_{i,f}}\,
  |\mathcal{M}|^2\,
  \delta\left((p_A+q)^2-W_X^2\right)\, \theta(E_X).
\end{equation}
Here $W_X^2 \equiv p_X^2$ is the invariant mass of the produced
hadronic state $X$ and $\epsilon_i$ is the incoming electron energy.
Introducing the DIS interaction vertex $\Gamma^\mu_{AX}$ between
initial nuclear ground sate $|\Psi_A\rangle$ and final state
$|\Phi_X\rangle$, one can represent the matrix element $\mathcal{M}$
of the scattering as the product of leptonic ($J^e_\mu$) and nuclear
($J^\mu_{AX}$) currents,
\begin{equation} \label{eq:inclmatrix}
 -i \mathcal{M}
= -{\frac{ie^2}{q^2}}
  J^e_\mu(k_f,s_{e'};k_i,s_e)\,
  J^\mu_{AX}(p_A,s_A;p_X,s_X),
\end{equation}
where
 $J^e_\mu = \bar{u}(k_f,s_{e'}) \gamma_\mu u(k_i,s_e)$
and
 $J^\mu_{AX}(p_A,s_A;p_X,s_X)
 = \langle \Phi(p_X,s_X) | \Gamma^\mu_{AX} | \Psi_A(p_A,s_A) \rangle$.
Here $s_e$ $(s_{e'})$, $s_X$ and $s_A$ are the spins of the incoming
(final) electron, hadronic state $X$ and nucleus $A$, respectively.
In terms of the currents, the differential cross section in
Eq.~(\ref{eq:crossA}) can be written in terms of leptonic
($L_{\mu\nu}$) and hadronic ($W^{\mu\nu}_A$) tensors,
\begin{equation} \label{eq:crossAfinal}
 \frac{d\sigma}{d\epsilon_f\Omega_f}
= \frac{\alpha^2}{Q^4} \frac{\epsilon_f}{\epsilon_i}
  L_{\mu\nu} W^{\mu\nu}_A,
\end{equation}
where $\alpha$ is the electromagnetic coupling.
The leptonic tensor is given by
\begin{equation}
 L_{\mu\nu}
= \frac{1}{2} \sum_{s_e,s_{e'}} J^{e \dagger}_\mu J^e_\nu
= 2 \left( k_{i \mu} k_{f \nu} + k_{f \mu} k_{i \nu}
	   + \frac{q^2}{2} g_{\mu\nu}
    \right),
\end{equation}
while the hadronic tensor can be formally written as
\begin{eqnarray} \label{eq:hadrtensor}
 W^{\mu\nu}_A
&=& \frac{1}{2M_A} \frac{1}{(2j_A+1)} \sum_X \sum_{s_A,s_X}
    J^{\mu \dagger}_{AX}(p_A,s_A;p_X,s_X)\,
    J^\nu_{AX}(p_A,s_A;p_X,s_X)		\nonumber\\
& & \hspace*{4.5cm}
\times\
    \delta\left( (p_A+q)^2-W_X^2 \right)\, \theta(E_X),
\end{eqnarray}
where $j_A$ is the total spin of the nucleus $A$.

\begin{figure}[t]
\begin{center}
\includegraphics[width=0.5\textwidth]{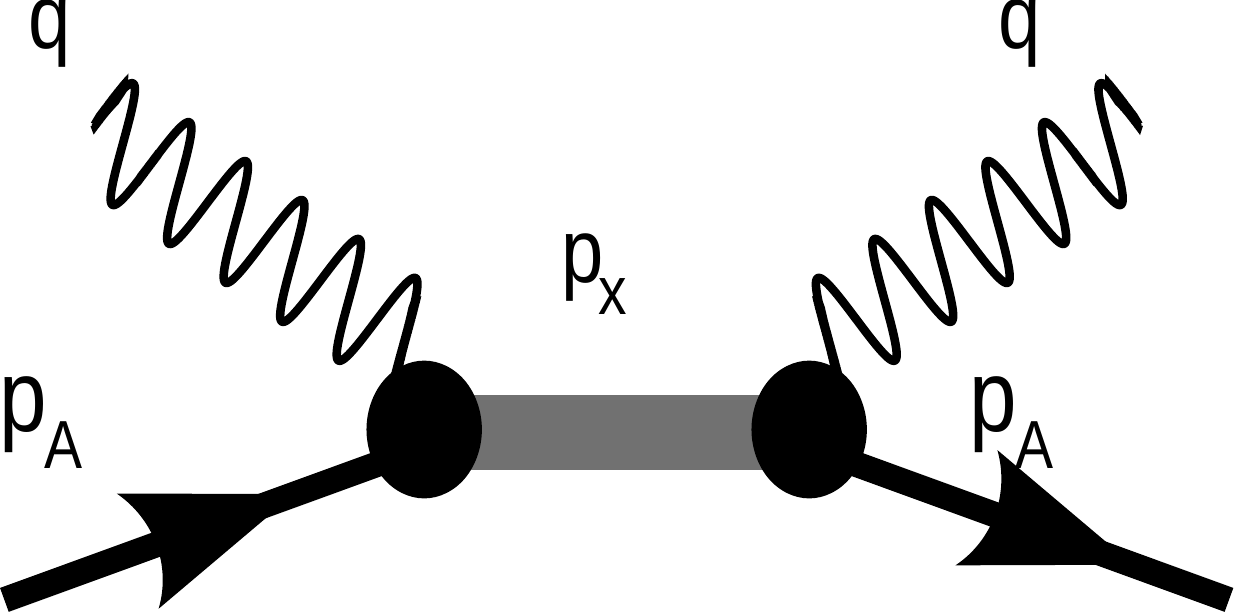}
\caption{Forward virtual Compton scattering amplitude from a
	nucleus $A$, with $q$ and $p_A$ the photon and target
	four-momenta, and $p_X$ the four-momentum of the
	produced state $X$.}
\label{fig:VCamp}
\end{center}
\end{figure}

Expressing the DIS differential cross section through the nuclear
hadronic tensor makes the application of the optical theorem rather
straightforward.  From the nuclear virtual Compton scattering amplitude
shown in Fig.~\ref{fig:VCamp} one observes that the imaginary part of
the intermediate state propagator, with the condition $E_X>0$,
corresponds to $\pi \delta\left( (p_A+q)^2 - W_X^2 \right)$.
This gives for the imaginary part of the Compton amplitude in the forward
direction ($t=0$),
\begin{equation}
 {\Im}m\, {\cal A}^{\mu\nu}_{\gamma^* A}(t=0)
 = \sum_X J^{\mu \dagger}_{AX}(p_A,s_A;p_X,s_X)\,
	  J^\nu_{AX}(p_A,s_A;p_X,s_X)\,
	  \pi \delta\left( (p_A+q)^2-W_X^2 \right)\, \theta(E_X),
\label{eq:Compton}
\end{equation}
where $s_A = s_{A^\prime}$ due to the forward elastic scattering
condition.
Comparing Eqs.~(\ref{eq:hadrtensor}) and (\ref{eq:Compton}) one
obtains the optical theorem relation between the nuclear hadronic
tensor and the forward nuclear Compton scattering amplitude,
\begin{equation}
 W^{\mu\nu}_A
= \frac{1}{2 \pi M_A} \frac{1}{(2j_A+1)}
  \sum_{s_A} {\Im}m\, {\cal A}^{\mu\nu}_{\gamma^* A}(t=0).
\label{eq:optical}
\end{equation}

\begin{figure}[tb]
\begin{center}
\includegraphics[width=0.7\textwidth]{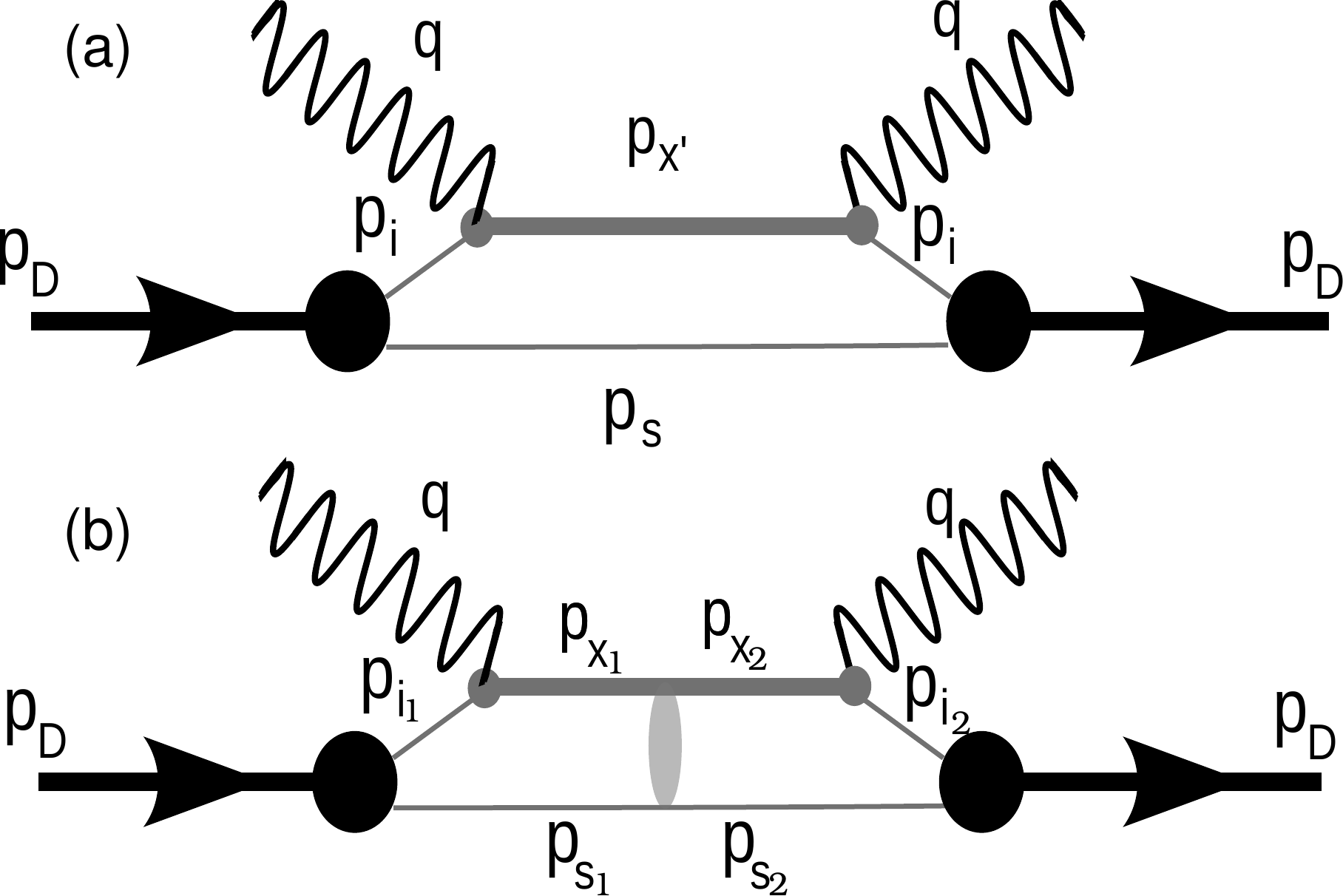}
\caption{Forward virtual Compton scattering amplitude for
	the deuteron, comprising of
	{\bf (a)} the Born diagram, and
	{\bf (b)} the rescattering contribution.
	The gray blob in the intermediate state represents the
	effective rescattering interaction of the hadronic debris
	($X_1$) and the spectator nucleon ($S_1$) to the final
	hadronic state ($X_2$) and nucleon ($S_2$).
	The deuteron momentum in the Born diagram is given by
	$p_D = p_i + p_s$, and in the FSI diagram by
	$p_D = p_{i_1} + p_{s_1} = p_{i_2} + p_{s_2}$.}
\label{fig:GEAamp}
\end{center}
\end{figure}

Although the above discussion holds for an arbitrary nucleus $A$,
we shall now focus on the specific case of the deuteron.
At large $Q^2$ and for deuteron internal momenta up to 700~MeV
(see e.g. Ref.~\cite{Frankfurt:2008zv}) one expects the virtual
photon scattering from the deuteron target to take place from an
individual nucleon bound in the nucleus.  This allows us to express
the Compton scatting amplitude as a sum of two terms, as illustrated
in Fig.~\ref{fig:GEAamp}.
The first (Born) term represents the propagation of the state $X'$
resulting from the $\gamma^*$--bound nucleon scattering, without
interacting with the spectator nucleon, Fig.~\ref{fig:GEAamp}(a).
The second (rescattering) term corresponds to the produced hadronic
state ($X_1$) interacting with the spectator nucleon ($S_1$) in the
intermediate state of the Compton scattering, Fig.~\ref{fig:GEAamp}(b).
The latter diagram is responsible for the FSI contribution to inclusive
DIS.

% ........................................................................
\subsection{Born term}
\label{subsec:born}

For the Born diagram of the forward Compton scattering amplitude
from the deuteron, following the prescription of the effective
Feynman diagram rules for inelastic scattering \cite{Cosyn:2010ux},
the plane-wave (pw) amplitude can be written as
\begin{multline}
 \mathcal{A}^{\mu\nu}_{\text{pw}}
= \sum_{N,X'} \int\!\frac{d^4p_s}{i(2\pi)^4}\,
  (\chi^{s_D})^\dagger\, \Gamma_{DNN}^\dagger\,
  \frac{\slashed{p}_i+m}{p_i^2-m^2+i\epsilon}\,
  \Gamma^{\mu\dagger}_{\gamma NX'}\,
  \frac{G(p_{X'})}{p_{X'}^2-m_{X'}^2+i\epsilon}	\\
\times
  \frac{\slashed{p}_s+m}{p_s^2-m^2+i\epsilon}\,
  \Gamma^{\nu}_{\gamma NX'}\,
  \frac{\slashed{p}_i+m }{p_i^2-m^2+i\epsilon}\,
  \Gamma_{DNN}\, \chi^{s_D},
\end{multline}
where $p_i$ is the four-momentum of the initial, off-shell nucleon
and $p_s$ is the four-momentum of the spectator nucleon, both with
mass $m$.  The sum runs over all possible intermediate states $X'$
and the proton and neutron contribution.  The inelastic intermediate
state is characterized by the momentum $p_{X'} = p_D + q - p_s$ and
mass $m_{X'}$.
The function $G(p_{X'})$ describes the Green function of the
intermediate state $X'$, $\Gamma^\mu_{\gamma NX'}$ is the
photon--nucleon vertex, and $\Gamma_{DNN}$ denotes the $DNN$
vertex function, with $\chi^{s_D}$ the deuteron spin wave function
for spin projection $s_D$.
In the virtual nucleon approximation (VNA) \cite{Cosyn:2010ux,
Sargsian:2009hf}, the loop integration over $dp_s^0$ is performed
by retaining only the positive energy, on-mass-shell contribution
of the spectator nucleon propagator,
\begin{equation} \label{eq:pole}
\int\frac{dp_s^0}{p^2_s - m^2 + i\epsilon}\
\longrightarrow\ -i\frac{\pi}{E_s},
\end{equation}
where $E_s = \sqrt{m^2 + \bm{p}_s^2}$ is the spectator nucleon energy.
Conservation of energy requires that the energy of the interacting,
off-shell nucleon is then equal to $E_i = M_D - E_s$.

It is convenient in the VNA to introduce the deuteron wave function
$\Psi_D^{s_D}(p_1,s_1;p_2,s_2)$ for the case of one nucleon ($p_1$)
being off-shell and one nucleon ($p_2$) on-shell \cite{Cosyn:2010ux,
Sargsian:2009hf, Gross:1991pm, Gribov:1968gs, Bertocchi:1972cj},
\begin{equation}
\label{eq:deuterondef}
\Psi_D^{s_D}(p_1,s_1;p_2,s_2)
= -\frac{\bar{u}(p_1,s_1)\, \bar{u}(p_2,s_2)\, \Gamma_{DNN}\, \chi^{s_D}}
	{(p_1^2-m^2) \sqrt{2} \sqrt{(2\pi)^3 2 E_2}},
\end{equation}
where $E_2 = \sqrt{m^2 + {\bm p}_2^2}$, and we take the masses
of the two nucleons to be equal, $m_1 = m_2 = m$.
This then allows the pw amplitude to be expressed as
\begin{multline} \label{eq:Afirst}
\mathcal{A}^{\mu\nu}_{\text{pw}}
= - 2 \sum_{N,X'} \sum_{s_i,s'_i,s_s}
  \int\!d^3{\bm p}_s\,
  \Psi_D^{s_D \dagger}(p_i,s'_i;p_s,s_s)\,
  \bar{u}(p_i,s'_i)\,
  \Gamma^{\mu \dagger}_{\gamma NX'}		\\
\times
  \frac{G(p_{X'})}{p_{X'}^2-m_{X'}^2+i\epsilon}\,
  \Gamma^\nu_{\gamma NX'}\,
  u(p_i,s_i)\,
  \Psi_D^{s_D}(p_i,s_i;p_s,s_s),
\end{multline}
where $s_i, s'_i$ are the spins of the off-shell nucleons
and $s_s$ is the spin of the spectator on-shell nucleon.
In general, the propagator of the inelastic intermediate state $X'$
can be represented as a sum of on-shell and off-shell contributions
\cite{Cosyn:2010ux},
\begin{multline} \label{eq:propX}
\frac{G(p_{X'})}{p_{X'}^2-m_{X'}^2+i\epsilon}
\approx
  \frac{\sum_{s_{X'}} |p_{X'},s_{X'}\rangle \langle p_{X'},s_{X'}|}
       {p_{X'}^2-m_{X'}^2+i\epsilon }	\\
= - i \pi \sum_{s_{X'}} |p_{X'},s_{X'}\rangle \langle p_{X'},s_{X'}|\,
  \delta(p_{X'}^2-m_{X'}^2)\, \theta(E_{X'})
+ \mathcal{P}
  \frac{\sum_{s_{X'}} |p_{X'},s_{X'}\rangle \langle p_{X'},s_{X'}|}
       {p_{X'}^2-m_{X'}^2},
\end{multline}
where in the off-shell term the symbol $\mathcal{P}$ denotes the
Cauchy principal value integration.
Since the imaginary part of the Compton scattering amplitude is
defined by the on-shell part of the inelastic state $X'$ propagator,
substituting Eq.~(\ref{eq:propX}) into Eq.~(\ref{eq:Afirst}) and
averaging over the deuteron polarizations, one obtains
\begin{multline} \label{eq:imAborn}
\frac{1}{3}\sum_{s_D}
 {\Im}m\, \mathcal{A}^{\mu\nu}_{\text{pw}}
= \pi \int\!d^3{\bm p}_s \sum_{N,X'} \sum_{s_i,s_{X'}}
  J^{\mu \dagger}_{\gamma NX'}(p_i,s_i;p_{X'},s_{X'})	\\
\times
  J^\nu_{\gamma NX'}(p_i,s_i;p_{X'},s_{X'})\,
  \delta(p_{X'}^2-m_{X'}^2)\, \theta(E_{X'})\,
  S(p_s),
\end{multline}
where the deuteron momentum distribution is defined as
\begin{eqnarray}
S(p_s)
&=& \frac{1}{3} \sum_{s_D,s_s,s_i}
    \Big| \Psi_D^{s_D}(p_i,s_i;p_s,s_s) \Big|^2.
\end{eqnarray}
Introducing the nucleon hadronic tensor in analogy with
Eq.~(\ref{eq:hadrtensor}),
\begin{equation} \label{eq:nucltensor}
W_N^{\mu\nu}
= \frac{1}{2m} \frac{1}{2}
  \sum_{X'} \sum_{s_i,s_{X'}}
  J^{\mu \dagger}_{\gamma NX'}(p_i,s_i;p_{X'},s_{X'})
  J^\nu_{\gamma NX'}(p_i,s_i;p_{X'},s_{X'})\,
  \delta(p_{X'}^2-m_{X'}^2)\, \theta(E_{X'}),
\end{equation}
and using it in Eq.~(\ref{eq:imAborn}), from the optical theorem
relation in Eq.~(\ref{eq:optical}) one obtains for the deuteron
hadronic tensor,
\begin{equation} \label{eq:WBorn}
W^{\mu\nu}_D
= \frac{2m}{M_D} \sum_N \int\!d^3{\bm p}_s\,
  W_N^{\mu\nu}\, S(p_s),
\end{equation}
where the sum is over the nucleons $N = p, n$.
This corresponds to the usual convolution model of inclusive
DIS from the deuteron \cite{Melnitchouk:1993nk, Kulagin:1994fz,
Sargsian:2001gu, Kulagin:2004ie}.

% ........................................................................
\subsection{Final-state interaction contribution}
\label{ssec:fsi}

Having outlined the basic derivation of the Born contribution to DIS
from the deuteron, we are now able to proceed with the calculation of
the FSI corrections to the inclusive deuteron structure functions.
The strategy will be to use the GEA to compute the rescattering
contribution in Fig.~\ref{fig:GEAamp}(b) to the forward Compton
scattering amplitude, and relate it to the inclusive DIS process
via the optical theorem, Eq.~(\ref{eq:optical}).  Note that in the
eikonal approximation the hadronic rescattering vertex is an
effective vertex related to the hadronic scattering amplitude.
Consequently, only the FSI amplitude in Fig.~\ref{fig:GEAamp}(b)
needs to be considered, where a sum over all intermediate states
$X_1$ and $X_2$ is taken, and higher order rescattering contributions
are included in the effective rescattering vertex.
Two conditions, however, must be satisfied for the GEA to be valid
in the calculation of the FSI contribution:
\begin{enumerate}
\item
The intermediate state can be characterized as an effective hadronic
state whose interaction with the spectator nucleon can have attributes
of the $hN$ interaction.  Such states comprise any intermediate state
resonances that can be generated at the first $\gamma^* N$ vertex in
Fig.~\ref{fig:GEAamp}(b).
\item
The produced state $X_1$ is energetic enough for the eikonal
approximation to be valid for $X_1 N$ rescattering, and it can be
described by a diffractive amplitude.
\end{enumerate}

\begin{figure}[t]
\begin{center}
\includegraphics[width=0.7\textwidth]{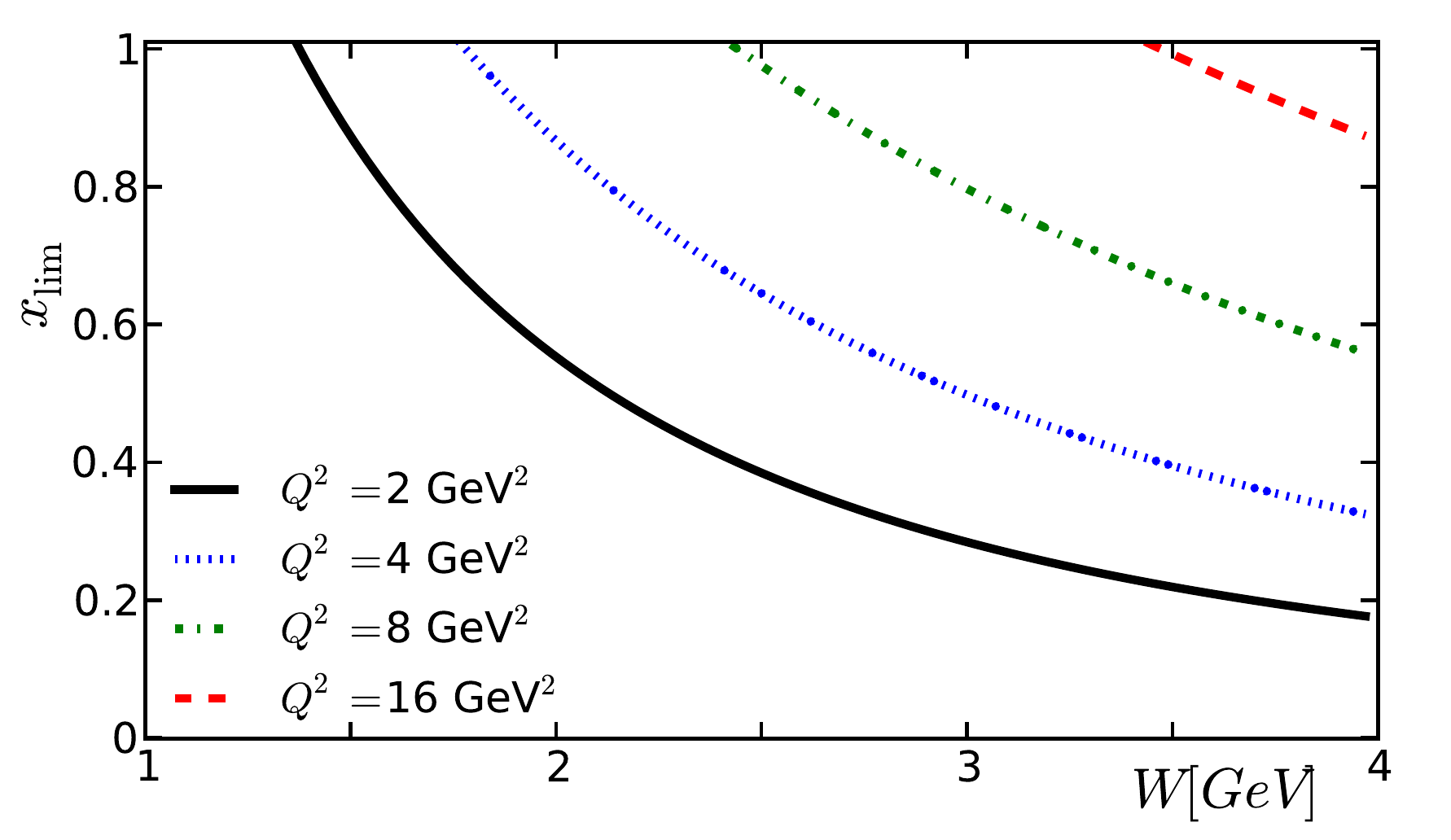}
\caption{(Color online)
	Maximum value of Bjorken $x$ allowed for the application
	of the eikonal approximation as a function of the invariant
	mass $W$ of the state $X'$.}
\label{fig:xlim}
\end{center}
\end{figure}

For the second condition we can use the empirical observation that in
hadron--nucleon scattering the eikonal approximation holds for hadron
momenta $\gtrsim 500$~MeV \cite{Pandharipande:1992zz, Laget:2004sm,
Ryckebusch:2003fc} in a frame where the nucleon is at rest.
This can be used to define the quantity $x_\text{lim}$ as the maximum
value of the Bjorken scaling variable $x = Q^2/2mq_0$ at a given $Q^2$
for which the state $X'$ in the rescattering system has a momentum of
at least 500~MeV.  This value of $x_\text{lim}$ can be found from the
relation
\begin{equation}
 s_{X'N} = (p_D + q)^2 = m^2 + m_{X'}^2 + 2 m E_{X'},
\end{equation}
where $E_{X'}$ is the energy of the state $X'$ in a frame where the
``spectator'' nucleon ($S_1$ in Fig.~\ref{fig:GEAamp}(b)) is at rest.
In Fig.~\ref{fig:xlim} we show $x_\text{lim}$ as a function of the
invariant mass $W$ of $X'$ taking part in the rescattering for several
values of $Q^2$.  For values of $Q^2$ up to 5~GeV$^2$ care should
be taken when $x>0.5$ for values of $W$ higher than 2~GeV entering
in the rescattering amplitude.

Assuming now that the chosen kinematics is appropriate for the
application of the eikonal approximation and that the intermediate
states $X$ are identified, one can apply the effective Feynman
diagram rules for the rescattering diagram of Fig.~\ref{fig:GEAamp}(b)
to obtain the FSI contribution to the virtual Compton amplitude,
\begin{multline} \label{eq:FSI}
\mathcal{A}^{\mu\nu}_{\text{FSI}}
= \sum_{N,X_1,X_2} \int\!
  \frac{d^4p_{s_1}}{i(2\pi)^4}
  \frac{d^4p_{s_2}}{i(2\pi)^4}
  (\chi^{s_D})^\dagger
  \Gamma_{DNN}^\dagger
  \frac{\slashed{p}_{i_2}\!+m}{p_{i_2}^2-m^2+i\epsilon}
  \Gamma^{\mu \dagger}_{\gamma N X_2}
  \frac{G(p_{X_2})}{(p_{X_2}^2-m_{X_2}^2+i\epsilon)}	\\
\times
  \frac{\slashed{p}_{s_2}\!+m}{p_{s_2}^2-m^2+i\epsilon}\,
  F_{NX_1,NX_2}\,
  \frac{G(p_{X_1})}{(p_{X_1}^2-m_{X_1}^2+i\epsilon)}
  \frac{\slashed{p}_{s_1}\!+m}{p_{s_1}^2-m^2+i\epsilon}
  \Gamma^{\nu}_{\gamma N X_1}
  \frac{\slashed{p}_{i_1}\!+m}{p_{i_1}^2-m^2+i\epsilon}
  \Gamma_{DNN} \chi^{s_D},
\end{multline}
where $F_{NX_1,NX_2}$ is the $N X_1 \to N X_2$ effective scattering
amplitude.  To evaluate the FSI amplitude, we first integrate over
$p^0_{s_1}$ and $p^0_{s_2}$ through the positive energy poles,
as in Eq.~(\ref{eq:pole}).  According to the VNA, we can then use
the on-mass-shell decomposition of the virtual nucleon propagators,
with their off-shell four-momenta $p_{i_1}$ and $p_{i_2}$ defined
through four-momentum conservation at the $DNN$ vertices, to introduce
the deuteron wave functions of Eq.~(\ref{eq:deuterondef}) into the
amplitude in Eq.~(\ref{eq:FSI}).
Finally, because of the large momenta involved in the propagators
of the intermediate states $X_j$ ($j=1,2$), an on-shell relation
for the Green function
$G(p_{X_j}) = \sum_{s_{X_j}}
	      |p_{X_j},s_{X_j} \rangle \langle p_{X_j},s_{X_j}|$
can be used, which allows the FSI amplitude to be written as
\begin{align} \label{eq:fullfsi}
 \mathcal{A}^{\mu\nu}_{\text{FSI}}
&= 2 (2\pi)^3 \sum_{N,X_1,X_2}
  \sum_{\substack{s_{i_1},s_{s_1},s_{X_1}\\s_{i_2},s_{s_2},s_{X_2}}}
  \int\frac{d^3{\bm p}_{s_1}}{(2\pi)^3}
      \frac{d^3{\bm p}_{s_2}}{(2\pi)^3}
  \frac{\Psi_D^{s_D \dagger}(p_{i_2},s_{i_2};p_{s_2},s_{s_2})}
       {\sqrt{2E_{s_2}}}
  \frac{J^{\mu \dagger}_{\gamma N X_2}(p_{i_2},s_{i_2};p_{X_2},s_{X_2})}
       {p_{X_2}^2-m_{X_2}^2+i\epsilon}		\nonumber\\
&\times
  \langle p_{X_2},s_{X_2};p_{s_2},s_{s_2} |F_{NX_1,NX_2}|
	  p_{X_1},s_{X_1};p_{s_1},s_{s_1}
  \rangle					\nonumber\\
&\times
  \frac{J^\nu_{\gamma NX_1}(p_{i_1},s_{i_1};p_{X_1},s_{X_1})}
       {p_{X_1}^2-m_{X_1}^2+i\epsilon}
  \frac{\Psi_D^{s_D}(p_{i_1},s_{i_1};p_{s_1},s_{s_1})}
       {\sqrt{2E_{s_1}}}.
\end{align}
In the following it will be useful to write the product of the
denominators of the intermediate inelastic state propagators as
\begin{align} \label{eq:propcancel}
 \frac{1}{\left(p_{X_2}^2-m_{X_2}^2+i\epsilon\right)}
&\frac{1}{\left(p_{X_1}^2-m_{X_1}^2+i\epsilon\right)}	\nonumber\\
&= \left(- i \pi \delta(p_{X_2}^2-m_{X_2}^2)
        + \frac{\mathcal{P}}{p_{X_2}^2-m_{X_2}^2}
  \right)
  \left(- i \pi \delta(p_{X_1}^2-m_{X_1}^2)
	+ \frac{\mathcal{P}}{p_{X_1}^2-m_{X_1}^2}
  \right)				\nonumber	\\
\hspace*{-4cm}
&= - \pi^2 \delta(p_{X_1}^2-m_{X_1}^2)
	  \delta(p_{X_2}^2-m_{X_2}^2)
  + \frac{\mathcal{P}}{p_{X_1}^2-m_{X_1}^2}
    \frac{\mathcal{P}}{p_{X_2}^2-m_{X_2}^2},
\end{align}
where the imaginary cross terms cancel exactly because of
energy-momentum conservation at the rescattering vertices in
the diagram of Fig.~\ref{fig:GEAamp}(b).  This decomposition
allows the amplitude in Eq.~(\ref{eq:fullfsi}) to be separated
into two terms containing on-shell and off-shell contributions
to the rescattering amplitude $F_{NX_1,NX_2}$.
It is also worth mentioning that the first part contains
half-off-shell inelastic electromagnetic currents, while
in the second part these are fully off-shell.
The first part of the decomposition (\ref{eq:propcancel})
yields the ``on-shell'' component of the FSI contribution
to the deuteron hadronic tensor,
\begin{align} \label{eq:generalFSIon}
W^{\mu\nu\text{(on)}}_{\text{FSI}}
&= -\frac{\pi(2\pi)^3}{3 M_D} \sum_{N,X_1,X_2}
  \sum_{\text{spins}}\,
  {\Im}m\, \int\frac{d^3{\bm p}_{s_1}}{(2\pi)^3}
               \frac{d^3{\bm p}_{s_2}}{(2\pi)^3}
  \frac{\Psi_D^{s_D \dagger}(p_{i_2},s_{i_2};p_{s_2},s_{s_2})
        \Psi_D^{s_D}(p_{i_1},s_{i_1};p_{s_1},s_{s_1})}
       {2\sqrt{E_{s_2}E_{s_1}}}			\nonumber\\
&\times
  \langle p_{X_2},s_{X_2};p_{s_2},s_{s_2}
  |F^{\text{(on)}}_{NX_1,NX_2}|
  p_{X_1},s_{X_1};p_{s_1},s_{s_1} \rangle
  J^{\mu \dagger}_{\gamma NX_2}(p_{i_2},s_{i_2};p_{X_2},s_{X_2})
						\nonumber\\
  &\times J^\nu_{\gamma NX_1}(p_{i_1},s_{i_1};p_{X_1},s_{X_1})		
  \delta(p_{X_1}^2-m_{X_1}^2)
  \delta(p_{X_2}^2-m_{X_2}^2),
\end{align}
where the spin summation includes the sum over $s_{i_1}$,
$s_{s_1}$, $s_{X_1}$, $s_{i_2}$, $s_{s_2}$, $s_{X_2}$ and $s_D$.
Finally, the second term in Eq.~(\ref{eq:propcancel}) enters with
opposite sign and represents the contribution of the off-shell
part of the FSI contribution to the deuteron hadronic tensor,
\begin{align} \label{eq:generalFSIoff}
W^{\mu\nu\text{(off)}}_{\text{FSI}}
&= \frac{(2\pi)^3}{3\pi M_D} \sum_{N,X_1,X_2}
  \sum_{\text{spins}}
  {\Im}m\ \int_{\mathcal{P}}
  \frac{d^3{\bm p}_{s_1}}{(2\pi)^3}
  \frac{d^3{\bm p}_{s_2}}{(2\pi)^3}
  \frac{\Psi_D^{s_D \dagger}(p_{i_2},s_{i_2};p_{s_2},s_{s_2})
        \Psi_D^{s_D}(p_{i_1},s_{i_1};p_{s_1},s_{s})}
       {2\sqrt{E_{s_2}E_{s_1}}}			\nonumber\\
&\times
  \langle p_{X_2},s_{X_2};p_{s_2},s_{s_2}
  |F^{\text{(off)}}_{NX_1,NX_2}|
  p_{X_1},s_{X_1};p_{s_1},s_{s_1} \rangle
  J^{\mu\dagger}_{\gamma NX_2}(p_{i_2},s_{i_2};p_{X_2},s_{X_2})
						\nonumber\\
  &\times J^\nu_{\gamma NX_1}(p_{i_1},s_{i_1};p_{X_1},s_{X_1})	
  \frac{1}{p_{X_1}^2-m_{X_1}^2}
  \frac{1}{p_{X_2}^2-m_{X_2}^2},
\end{align}
where $\int_{\mathcal{P}}$ indicates Cauchy principal value integration.
The total FSI contribution to the deuteron hadronic tensor is then
given by the sum
$W^{\mu\nu}_{\text{FSI}}
= W^{\mu\nu\text{(on)}}_{\text{FSI}}
+ W^{\mu\nu\text{(off)}}_{\text{FSI}}$.
The results in Eqs.~(\ref{eq:generalFSIon}) and (\ref{eq:generalFSIoff})
form the general expression for the FSI contribution to the deuteron
hadronic tensor within the eikonal approximation.  To evaluate these
expressions in practice requires modeling of the matrix elements of
$F_{NX_1,NX_2}$ and a truncation of the set of states comprising the
inelastic intermediate states in Fig.~\ref{fig:GEAamp}, which we turn
to in the next section.

%%%%%%%%%%%%%%%%%%%%%%%%%%%%%%%%%%%%%%%%%%%%%%%%%%%%%%%%%%%%%%%%%%%%%%%%%
\section{Factorized Effective Resonance Model} \label{sec:estimates}

Considerable experience has been developed over the years with
the computation of the Born approximation contribution
[Eq.~(\ref{eq:WBorn})] to the deuteron hadronic tensor,
especially since the observation of the nuclear EMC effect,
or deviation from unity of the ratio of nuclear to deuterium
structure functions, some three decades ago \cite{Aubert:1983xm}.
For the deuteron, the input required in these calculations is the
nucleon momentum distribution in the deuteron and the bound nucleon
structure functions in the nucleon hadronic tensor $W_N^{\mu\nu}$.
In the VNA the latter is approximated by the on-shell hadronic tensor,
with the off-shell nucleon momentum defined through the on-shell
momentum of the deuteron and spectator nucleon.  Some attempts have
been made, however, to account for the possible effects on the bound
nucleon structure of nuclear medium modification, or off-shell
corrections (see e.g. \cite{Frankfurt:1985cv, Frankfurt:1988nt,
Gross:1991pi, Melnitchouk:1993nk, Kulagin:1994fz, Frank:1995pv,
Kulagin:2004ie, CJ11}.

Much less is understood about the FSI corrections
[Eqs.~(\ref{eq:generalFSIon}) and (\ref{eq:generalFSIoff})],
on the other hand, which require knowledge of the inelastic
currents and rescattering amplitude for the interaction between
the hadronic debris of the (off-shell) scattered nucleon and the
(on-shell) spectator nucleon.  The complexity of describing this
interaction is formidable, however, and as an exploratory attempt
to estimate the FSI effects numerically we use several additional
assumptions in order to make the calculation feasible.

% ........................................................................
\subsection{Factorized approximation for FSI}
\label{subsec:fac}

In the factorized approximation [also referred to as the distorted
wave impulse approximation (DWIA)], the inelastic electromagnetic
currents in both Eqs.~(\ref{eq:generalFSIon}) and
(\ref{eq:generalFSIoff}) are diagonalized by factoring out the
current $J^{\mu\dagger}_{\gamma NX_2}$ from the $d^3p_{s_2}$
integration evaluating it at 
$p_{i_2} = p_{i_1}$,  $s_{i_2} = s_{i_1}$,
$p_{X_2} = p_{X_1}$ and $s_{X_2} = s_{X_1}$
(and hence $m_{X_1} \approx m_{X_2}$).
This approximation is well known in quasi-elastic nuclear processes,
where it allows meaningful bound nucleon electromagnetic structure
functions to be identified when considering FSI effects
(see e.g. Ref.~\cite{Boffi:1993gs}).  In the following, we proceed
with the DWIA derivations of on-shell [Eq.~(\ref{eq:generalFSIon})]
and off-shell [Eq.~(\ref{eq:generalFSIoff})] hadronic tensors
separately.
We note also that the factorization approximation breaks the explicit
symmetry of Eqs.~(\ref{eq:generalFSIon}) and (\ref{eq:generalFSIoff})
with respect to the ``1'' and ``2'' indices.  However, in numerical
calculations this symmetry can be restored by taking the average of
results for the above factorization and one following from the
factorization of the $J^\nu_{\gamma NX_1}$ current from the
$d^3p_{s_2}$ integration.

For the on-shell hadronic tensor we  first express the $X_1 N \to X_2 N$
invariant scattering amplitude through the diffractive amplitude,
imposing the condition for helicity conservation in the form
\begin{multline} \label{eq:amp}
\langle p_{X_2},s_{X_2};p_{s_2},s_{s_2} |F_{NX_1,NX_2}(s_{XN},t_{XN})|
	p_{X_1},s_{X_1};p_{s_1},s_{s_1}
\rangle						\\
= \eta(s_{XN},m_{X_1})\,
  \langle p_{X_2},p_{s_2}|f_{NX_1,NX_2}(s_{XN},t_{XN})|
	  p_{X_1},p_{s_1}
  \rangle\,
  \delta_{s_{s_1},s_{s_2}} \delta_{s_{X_1},s_{X_2}}\,,
\end{multline}
where
$\eta(s_{XN},m_{X_1})
= \sqrt{[s_{XN}-(m-m_{X_1})^2][s_{XN}-(m+m_{X_1})^2]}$,
with the Mandelstam variables
$s_{XN} = (p_{X_1}+p_{s_1})^2=(p_{X_2}+p_{s_2})^2$ and
$t_{XN} = (p_{X_1}-p_{X_2})^2=(p_{s_1}-p_{s_2})^2$.
The amplitude $f_{N X_1, N X_2}$ is defined in the eikonal approximation,
\begin{equation} \label{eq:diffractive}
f_{NX_1,NX_2}(t_{XN})
= \sigma_{\text{tot}}(i+\epsilon)\, \exp(\beta t_{XN}/2),
\end{equation}
where $\sigma_{\text{tot}}$ represents the total cross section of the
scattering of the produced $X'$ system from the spectator nucleon,
$\beta$ is the slope factor, and $\epsilon$ represents the ratio of
the real to imaginary parts of the amplitude.

The approximation of helicity conservation allows the spin summations
in the electromagnetic current to be factorized from the final-state
interaction.  Combining then the factorized current
$J^{\mu\dagger}_{\gamma NX_1}$ with
$J^{\nu}_{\gamma NX_1}$ and $\delta(p_{X_1}^2-m_{X_1}^2)$,
one obtains the hadronic tensor of the bound nucleon
$W^{\mu\nu}_N(p_{i_1},Q^2,m_{X_1})$ using Eq.~(\ref{eq:nucltensor}).
In the DWIA this tensor is defined by the same kinematical conditions
as the tensor in the Born term of Eq.~(\ref{eq:WBorn}), with the
invariant mass $m_{X_1}^2 = (p_D-p_{s_1}+q)^2$ determined through
the integration over the spectator momentum $\bm p_{s_1}$,
and the momentum fraction of the initial struck nucleon is
\begin{equation}
x_1 = \frac{Q^2}{2 p_{i_1} \cdot q}
    = \frac{m\, x}{M_D - E_{s_1} + p_{s_1,z} |\bm q|/q^0}
    < 1.
\label{eq:x1}
\end{equation}
Note that even though the kinematics entering in
$W^{\mu\nu}_N(p_{i_1},Q^2,m_{X_1})$ are the same as for the 
Born contribution, we still retain the subscripts ``1'' in order
to distinguish these from the kinematics after rescattering.

For the rescattering part of Eq.~(\ref{eq:generalFSIon}),
the integration region is restricted by the condition on
the momentum fraction of the final nucleon $p_{i_2}$,
\begin{equation}
x_2 = \frac{Q^2}{2 p_{i_2} \cdot q}
    < 1.
\label{x_2on}
\end{equation}
The $p_{s_2,z}$ integration in Eq.~(\ref{eq:generalFSIon}) is taken
using the remaining $\delta$-function,
\begin{equation} \label{eq:psz}
\delta(p_{X_2}^2 - m_{X_2}^2)
= \frac{1}{2}
  \left| |{\bm q}|
	 - (M_D+q^0)
	   \frac{\widetilde{p}^{X_2}_{s_2,z}}{\widetilde{E}_{s_2}}
  \right|^{-1}
  \delta(p_{s_2,z}-\widetilde{p}^{X_2}_{s_2,z}),
\end{equation}
where 
$\widetilde{E}_{s_2}
= \sqrt{m^2 + {\bm p}_{s_2,\perp}^2 + (\widetilde{p}^{X_2}_{s_2,z})^2}$,
and $\widetilde{p}^{X_2}_{s_2,z}$ is obtained from the solution of
\begin{equation} \label{eq:pssolve}
  2 |{\bm q}|\, \widetilde{p}^{X_2}_{s_2,z}
- 2 (M_D+q^0) \widetilde{E}_{s_2}
= m_{X_2}^2 - M_D^2 + Q^2 - 2 M_D q^0 - m^2.
\end{equation}
The FSI part of the structure function will then be defined by the
integration of the transverse component $\bm p_{s_2,\perp}$ of the
spectator momentum.
Combining the factorized inelastic nucleon structure function with
the on-shell part of the FSI term of Eq.~(\ref{eq:generalFSIon}),
one finally obtains for the on-shell FSI contribution
\begin{align} \label{eq:fsifinalon}
W^{\mu\nu\text{(on)}}_{\text{FSI}}
&= - \sum_N \frac{2m}{M_D}
 \int\!d^3{\bm p}_{s_1}
 \frac{W_N^{\mu\nu}(p_{i_1},q,m_{X_1})}{8\sqrt{E_{s_1}}}
 \frac{1}{3} \sum_{X_2} \sum_{s_i,s_s,s_D}
 \int\!\frac{d^2{\bm p}_{s_2,\perp}}{(2\pi)^2}	\nonumber\\
&\times
 {\Im}m
 \Big\{
 \frac{\eta(s_{XN},m_{X_1})f^{\text{(on)}}_{N\{X_1\},NX_2}(t_{XN})}
      {\Big||{\bm q}|-(M_D+q^0)\,
		\widetilde{p}^{X_2}_{s_2,z}/\widetilde{E}_{s_2}
       \Big| \sqrt{\widetilde{E}_{s_2}}}
 C(p_{i_1}, \widetilde{p}^{X_2}_{i_2},q)	\nonumber\\
&\times
 \Psi_D^{s_D \dagger}
   (\widetilde{p}^{X_2}_{i_2},s_i;\widetilde{p}^{X_2}_{s_2},s_s)
 \Psi_D^{s_D}
   (p_{i_1},s_i;p_{s_1},s_s)
\Big\},
\end{align}
where the four-vectors
$\widetilde{p}^{X_2}_{s_2}
= (\widetilde{E}_{s_2};
   {\bm p}_{s_2,\perp},
   \widetilde{p}^{X_2}_{s_2,z})$
and
$\widetilde{p}^{X_2}_{i_2}
= p_D - \widetilde{p}^{X_2}_{s_2}$,
with $\{X_1\}$ denoting intermediate states that obey
$p^2_{X_1} = m^2_{X_1}$, and we use the shorthand notation
$s_i \equiv s_{i_1} = s_{i_2}$ and $s_s \equiv s_{s_1} = s_{s_2}$.
The symmetrization factor $C(p_{i_1},\widetilde{p}^{X_2}_{i_2},q)$
accounts for the choice of momenta at which the currents are
evaluated in the factorization approximation
(see Sec.~\ref{sec:results} below).

Using the solution of Eq.~(\ref{eq:pssolve}) one obtains the
more explicit expression for the momentum fraction $x_2$ entering
in the FSI part of Eq.~(\ref{eq:generalFSIon}),
\begin{equation}
x_2 = \frac{1}{1 + (m_{X_2}^2 - (\tilde p^{X_2}_{i_2})^2)/Q^2}
\approx
\frac{1}{1 + (m_{X_2}^2 - m^2)/Q^2}.
\end{equation}
The important feature of this expression is that it shows that
for any fixed value of $m_{X_2}$, the FSI terms are suppressed
kinematically in the Bjorken limit, where $x_2 \to 1$.
 
\medskip

For the off-shell part of the FSI amplitude of
Eq.~(\ref{eq:generalFSIoff}) one performs a similar factorization
of the electromagnetic currents, using helicity conservation in
the rescattering amplitude [as in Eq.~(\ref{eq:amp})].
The principal value integrations can be performed in
Eq.~(\ref{eq:generalFSIoff}) employing the method outlined in
Ref.~\cite{Cosyn:2010ux}, and using the analytic structure of the
deuteron wave function one obtains
\begin{equation}\label{eq:wave1}
\int_{\cal P}dp_{s,z}
\frac{\Psi_D^{s_D}(p_i,s_i;p_s,s_s)}
     {p_{s,z}-\widetilde{p}^X_{s,z}}
= -\pi \widetilde{p}^{X}_{s,z}
  \widetilde{\Psi}_D^{s_D}(\widetilde{p}^X_i,s_i;\widetilde{p}^{X}_s,s_s),
\end{equation}
with $\widetilde{\Psi}_D^{s_D}$ representing the distorted wave function
of the deuteron whose explicit form is given in Ref.~\cite{Cosyn:2010ux}.
After performing the above factorization and the principal value
integrations over $dp^z_{s_1}$ and $d p^z_{s_2}$ for the off-shell part
of the FSI of Eq.~(\ref{eq:generalFSIoff}), one finds
\begin{align}
\label{eq:fsioff_interm}
W^{\mu\nu\text{(off)}}_{\text{FSI}}
&=\frac{1}{6M_D} \sum_{N,X_1,X_2} \sum_{s_i,s_s,s_D} 
  \int d^2{\bm p}_{s_{1,\perp}}\widetilde{p}^{X_1}_{s_{1,z}}
  \frac{J^{\mu\dagger}_{\gamma NX_1}
  (\widetilde{p}^{X_1}_{i_1},s_{i};p_{X_1},s_{X})
  J^\nu_{\gamma NX_1}(\widetilde{p}^{X_1}_{i_1},s_{i};p_{X_1},s_{X})}
       {8\Big||{\bm q}|-(M_D+q^0)\,
		\widetilde{p}^{X_1}_{s_1,z}/\widetilde{E}_{s_1}
	 \Big|\sqrt{\widetilde{E}_{s_1}}} \nonumber\\
&\times
 \int
 \frac{d^2{\bm p}_{s_{2,\perp}}}{(2\pi)^2}\widetilde{p}^{X_2}_{s_{2,z}}\,
 {\Im}m
 \Big\{
   \frac{\eta(s_{XN},m_{X_1}) f^{\text{(off)}}_{NX_1,NX_2}(t_{XN})}
	{\Big| |{\bm q}|-(M_D+q^0)\,
	       \widetilde{p}^{X_2}_{s_{2,z}}/\widetilde{E}_{s_2}
	 \Big| \sqrt{\widetilde{E}_{s_2}}}	\nonumber\\
&\times
 \widetilde{\Psi}_D^{s_D\dagger}
   (\widetilde{p}^{X_1}_{i_1},s_i;\widetilde{p}^{X_1}_{s_1},s_s)
 \widetilde{\Psi}_D^{s_D}
   (\widetilde{p}^{X_2}_{i_2},s_i;\widetilde{p}^{X_2}_{s_2},s_s)
 \Big\}.
\end{align}
However, in order to introduce an off-shell nucleon hadronic tensor
$W_N^{\mu\nu (\text{off})}$ one needs to add an additional integral,
\begin{equation}
 1 = \int\!dW^2\, \delta(p_{X_1}^2-W^2),
\label{eq:wof}
\end{equation}
that allows the inelastic electromagnetic currents to be combined
to form the hadronic tensor of the nucleon in the form of
Eq.~(\ref{eq:nucltensor}).  Finally, we obtain for the off-shell
tensor
\begin{align}\label{eq:fsifinaloff}
W^{\mu\nu\text{(off)}}_{\text{FSI}}
&=\sum_{N,X_2}
  \frac{2m}{M_D}\int dW\, W
  \int d^2{\bm p}_{s_{1,\perp}}\widetilde{p}^{X_2}_{s_{1,z}}
  \frac{W_N^{\mu\nu (\text{off})}(\tilde{p}^{X_2}_{i_1},q,W)}
       {8\Big||{\bm q}|-(M_D+q^0)\,
		\widetilde{p}^{X_2}_{s_1,z}/\widetilde{E}_{s_1}
	 \Big|\sqrt{\widetilde{E}_{s_1}}}		\nonumber\\
&\times
 \frac{1}{3}  \sum_{s_i,s_s,s_D} \int
 \frac{d^2{\bm p}_{s_{2,\perp}}}{(2\pi)^2}\tilde{p}^{X_2}_{s_{2,z}}\,
 {\Im}m
 \Big\{
   \frac{\eta(s_{XN},W) f^{\text{(off)}}_{N\{X_1\}_W,NX_2}(t_{XN})}
	{\Big| |{\bm q}|-(M_D+q^0)\,
	       \widetilde{p}^{X_2}_{s_{2,z}}/\widetilde{E}_{s_2}
	 \Big| \sqrt{\widetilde{E}_{s_2}}}
  C(\widetilde{p}^{X_2}_{i_1},\widetilde{p}^{X_2}_{i_2},q) \nonumber\\
&\times
  \widetilde{\Psi}_D^{s_D\dagger}
    (\widetilde{p}^{X_2}_{i_1},s_i;\widetilde{p}^{X_2}_{s_1},s_s)
  \widetilde{\Psi}_D^{s_D}
    (\widetilde{p}^{X_2}_{i_2},s_i;\widetilde{p}^{X_2}_{s_2},s_s)
  \Big\},
\end{align}
where $\{X_1\}_W$ here are intermediate states with $p^2_{X_1} = W^2$.
Because the summation over $X_1$ is absorbed in the definition of the
nucleon hadronic tensor in Eq.~(\ref{eq:nucltensor}), the longitudinal
component of the initial spectator nucleon momentum
$\widetilde{p}^{X_2}_{s_1,z}$ is evaluated at $m_{X_2}$ in
Eq.~(\ref{eq:pssolve}), since this will minimize $t_{XN}$
($\widetilde{p}^{X_2}_{s_1,z} \approx \widetilde{p}^{X_2}_{s_2,z}$).
This maximizes the eikonal rescattering amplitude
$f^{\text{(off)}}_{N\{X_1\}_W,NX_2}(t_{XN})$, allowing us to
provide an estimate for the maximum possible effect of FSI.

Notice that in Eq.~(\ref{eq:fsifinaloff}) the nucleon hadronic tensor
and rescattering amplitude are computed at the Bjorken scaling variables
$x_1$ and $x_2$, respectively.  They can be calculated using the identity
in Eq.~(\ref{eq:wof}) and the property of the principal value integration
in Eq.~(\ref{eq:wave1}) which defines the longitudinal component
$p_{i_2,z}$ through energy-momentum conservation, Eq.~(\ref{eq:pssolve}).
This yields the approximate expressions
\begin{equation} \label{eq:tildex}
x_1 \approx \frac{1}{1 + (W^2 - m^2)/Q^2}\ \ \ \ \mbox{and}  \ \ \
x_2 \approx \frac{1}{1 + (m_{X_2}^2 - m^2)/Q^2},
\end{equation}
where $W$ is the mass over which the bound nucleon hadronic tensor is
integrated, and $m_{X_2}$ is the mass of the intermediate inelastic
state produced in the rescattering.  Similar to the on-shell case,
here too one finds that at large $Q^2$ both $x_1$ and $x_2 \to 1$,
suppressing the FSI contribution.  We note that the disappearance of
the FSIs in the Bjorken limit is independent of the factorization
approximation, since in general the values of $x_1$ and $x_2$ are
defined by $m_{X_1}$ and $m_{X_2}$ [similar to Eq.~(\ref{eq:tildex})].
The observation that in the Bjorken limit the FSIs disappear follows
therefore from the general property of the eikonal approximation
which assumes a rescattering between effective hadronic states with
masses $m_{X_1}$ and $m_{X_2}$.

\begin{figure}[t]
\begin{center}
\includegraphics[width=0.7\textwidth]{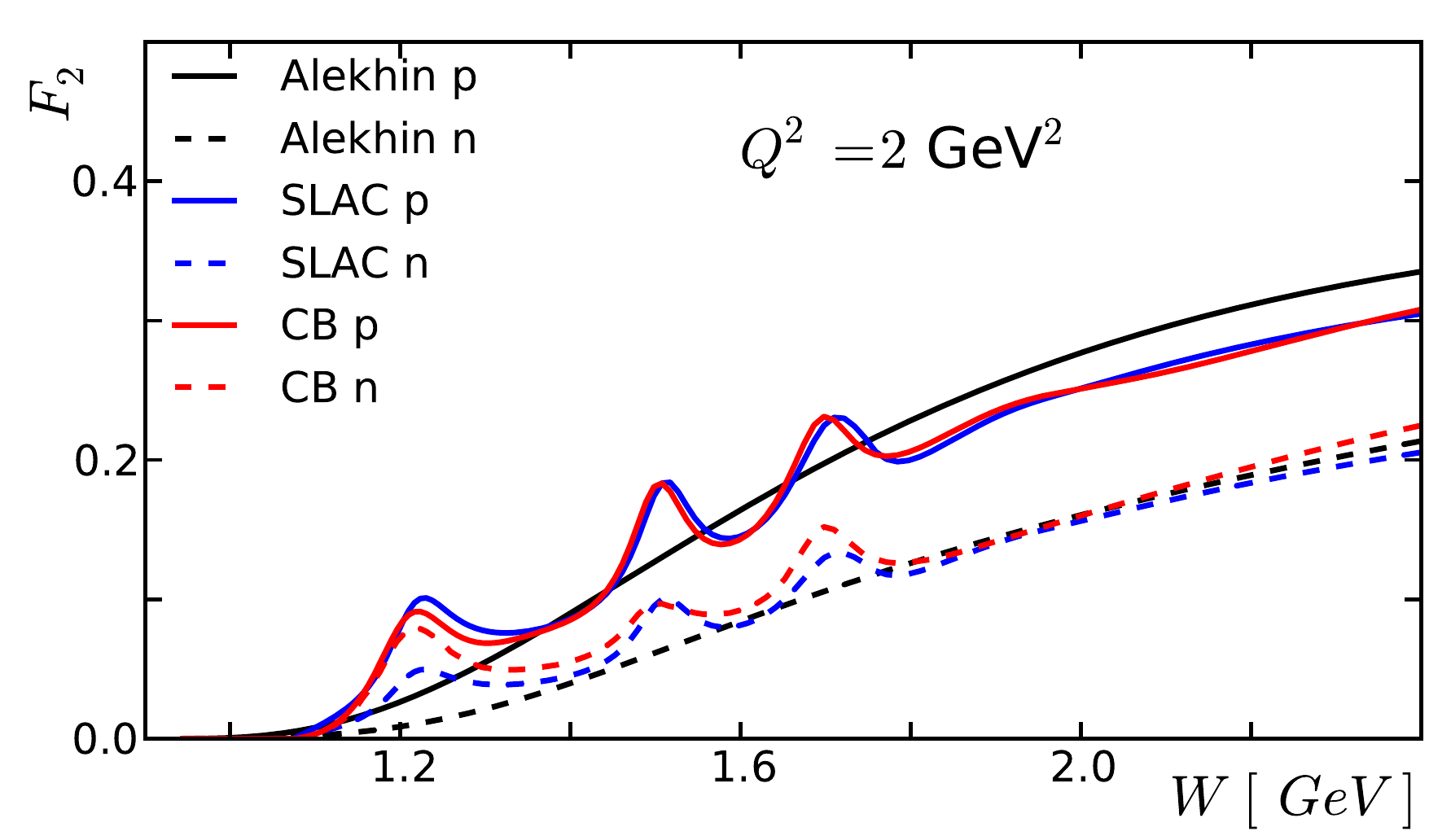}
\caption{(Color online)
	$F_2$ proton (solid lines) and neutron (dashed lines)
	structure functions at $Q^2=2$~GeV$^2$ for the Alekhin
	leading twist \cite{Alekhin:2002fv} (black), SLAC
	\cite{PhysRevD.20.1471} (blue), and Christy \& Bosted (CB)
	\cite{Christy:2007ve} (red) parametrizations of $F_2$.}
\label{fig:f2nucl}       % Give a unique label
\end{center}
\end{figure}

% ........................................................................
\subsection{Constraints from SIDIS}
\label{subsec:threeres}

In the earlier studies of FSI effects in SIDIS \cite{Cosyn:2010ux},
the mass produced in the intermediate state by the virtual photon
scattering from the moving nucleon was fixed by the kinematics of
the detected spectator nucleon.  In the inclusive DIS case, however,
the sum must be taken in Eqs.~(\ref{eq:fsifinalon}) and
(\ref{eq:fsifinaloff}) over all possible intermediate inelastic
states produced at the $\gamma^* N X$ vertex that can rescatter
from the spectator nucleon.

Inclusion of the complete set of states in Fig.~\ref{fig:GEAamp}(b)
that can contribute to the rescattering is of course not feasible,
and in practice a truncation of the spectrum is necessary, which
inevitably introduces an element of model dependence into the
calculation.
From measurements of nucleon inelastic structure functions at
intermediate values of $Q^2$, three clear resonance structures
are seen to dominate the spectrum in the low-$W$ region, as
illustrated in Fig.~\ref{fig:f2nucl}, centered at masses
$W_{\text{res}} = 1.232$~GeV, 1.5~GeV and 1.75~GeV.
In the present model we do not account for any relative phases
between the amplitudes for the different resonance contributions,
so that our estimate provides an upper bound on the size of the
FSI effects in this kinematic region.
From Fig.~\ref{fig:xlim} it is clear that for
$W_{\text{res}} < 1.75$~GeV the value of $x_\text{lim}$ remains
unity for $Q^2$ as low as 4~GeV$^2$, and is still reasonably high
($x_\text{lim} \gtrsim 0.7$) for $Q^2 = 2$~GeV$^2$, suggesting that
the eikonal approximation for the three-resonance model should be
valid down to these $Q^2$ values.

For larger invariant masses, above the resonance region, the analysis
\cite{Cosyn:2010ux} of deuteron SIDIS data \cite{Klimenko:2005zz}
suggests that FSIs yield sizeable contributions at the highest measured
$W$ bins ($W = 2.0$ and 2.4~GeV), and should therefore also be included
in our numerical estimates.  As discussed in Sec.~\ref{sec:intro},
however, for large invariant masses the phase space for the
intermediate state becomes unrestricted and the closure relation is
expected to hold in inclusive scattering.  To include contributions
from the $W \gtrsim 2$~GeV region, where no clear resonance structure
is visible, we consider specific widths in $W$ and fold the FSI
contributions of Eqs.~(\ref{eq:fsifinalon}) and (\ref{eq:fsifinaloff})
with a distribution in $W$, normalized to the values extracted from
the analysis of the SIDIS data \cite{Cosyn:2010ux}.  This simple
approximation allows the effects of FSIs to be estimated at small $x$
values, in addition to those at large $x$ which are determined by the
resonance contributions.  Note, however, that care must be taken when
including contributions with high $W$ ($W \gtrsim 3.5$~GeV), as at low
$Q^2$ values the intermediate states have momenta below the limit where
the eikonal approximation is known to be reliable.

The overall FSI contribution to the deuteron structure function
is therefore obtained by computing the total FSI hadronic tensor,
\begin{equation} \label{eq:sumres}
W^{\mu\nu\text{(tot)}}_{\text{FSI}}
= \int d m_{X_2} \left(\sum_{\text{res}}
  \delta(m_{X_2}-W_\text{res})+ \rho(m_{X_2})\right)
  \left( W^{\mu\nu\text{(on)}}_{\text{FSI}}
       + W^{\mu\nu\text{(off)}}_{\text{FSI}}
  \right),
\end{equation}
where the sum runs over the above-mentioned resonances and
$\rho(m_{X_2})$ denotes the spectral function representing
the DIS continuum region ($W>2$~GeV).
In the following section we present numerical estimates for
the FSI effects on the inclusive deuteron structure functions.

%%%%%%%%%%%%%%%%%%%%%%%%%%%%%%%%%%%%%%%%%%%%%%%%%%%%%%%%%%%%%%%%%%%%%%%%%
\section{Results} \label{sec:results}

In this section we examine the effects of including the FSI contribution
to the inclusive DIS cross section of the deuteron using the approach
outlined in Sec.~\ref{sec:estimates}.  We present calculations for the
inclusive structure function $F_2^D$, which is related to the
semi-inclusive structure functions $F_L^D$ and $F_T^D$ discussed
in Ref.~\cite{Cosyn:2010ux},
\begin{equation} \label{eq:f2ddef}
F_2^D
= \sum\limits_{N}
  \int\!\frac{d^3{\bm p}_s}{(2\pi)^2 2E_s}
  \left[ F_L^D(Q^2,\tilde x,{\bm p}_s)
%	+ \frac{Q^2}{2{\bm q}^2}}{\frac{q^0}{m}
	+ \frac{x}{\gamma^2}\,
	  F_T^D(Q^2,\tilde x,{\bm p}_s)
  \right],
\end{equation}
where $\gamma^2 \equiv {\bm q}^2/q_0^2 = 1 + 4 x^2 m^2/Q^2$, and
the variable $\tilde x$ is defined according to Eq.~(\ref{eq:x1}).
Unless noted otherwise, we use the SLAC parametrization
\cite{PhysRevD.20.1471} for the proton and neutron structure
functions, which covers a large range of $W$ and $Q^2$,
including the nucleon resonance structure.
To parametrize the $W$ and $Q^2$ dependence of the rescattering
parameters entering in Eq.~(\ref{eq:diffractive}), we use the
results of the analysis \cite{Cosyn:2010ux} of the semi-inclusive
DIS data from Ref.~\cite{Klimenko:2005zz}.
For the ratio of the real to imaginary parts of the cross
section and the slope factor, we take fixed values
$\epsilon = -0.2$ and $\beta=8~\text{GeV}^2$, respectively.
In the region of $W$ between the $\Delta(1232)$ resonance and the
highest mass accessed in the experiment \cite{Klimenko:2005zz},
$W = W_0 \equiv 2.4$~GeV, the extracted cross section rises
linearly with $W$ while falling with $Q^2$,
\begin{equation}
\sigma_{\text{tot}}(W,Q^2) = \frac{a+b\, W}{Q^2},
\end{equation}
with the constants $a = 25.3~\text{mb}$ and $b=53~\text{mb/GeV}$
fitted to the analysis of Ref.~\cite{Cosyn:2010ux}.  At higher $W$,
where the DIS cross section exhibits scaling, we take
$\sigma_{\text{tot}}(W > W_0,Q^2) = \sigma_{\text{tot}}(W_0,Q^2)$.
To further account for the off-shell nature of the inelastic
intermediate states in the hadronic currents and rescattering
amplitude in Eq.~(\ref{eq:fsifinaloff}), we introduce a suppression
factor \cite{Sargsian:2009hf} for the off-shell amplitude,
\begin{equation}
f^{\text{(off)}}_{N X_1,N X_2}(t_{XN})
= f^{\text{(on)}}_{N X_1,N X_2}(t_{XN})\,
  \exp\left( \beta\, |W^2-m^2_{X_2}|/2 \right),
\label{eq:off-ampl}
\end{equation}
where $W$ is the invariant mass that enters in the nucleon
hadronic tensor.  For the nonrelativistic deuteron wave function
we use the parametrization based on the Paris $NN$ potential
\cite{Lacombe:1980dr}, but also compare the results with the
CD-Bonn \cite{Machleidt:2000ge} and WCJ-1 \cite{Gross:2008ps}
wave functions.
The symmetrization factor $C(p_{i_1},p_{i_2},q)$ introduced in
Eqs.~(\ref{eq:fsifinalon}) and (\ref{eq:fsifinaloff}) is taken as
\begin{equation} 
C(p_{i_1},p_{i_2},q)
= \sqrt{\frac{F_2^N(x_2,Q^2)}{F_2^N(x_1,Q^2)}}\,,
\end{equation}
where $F_2^N$ is the nucleon structure function, and ensures that the
kinematical constraints $x_1, x_2 < 1$ are taken into account.

\begin{figure}[t]
\begin{center}
\includegraphics[width=0.5\textwidth]{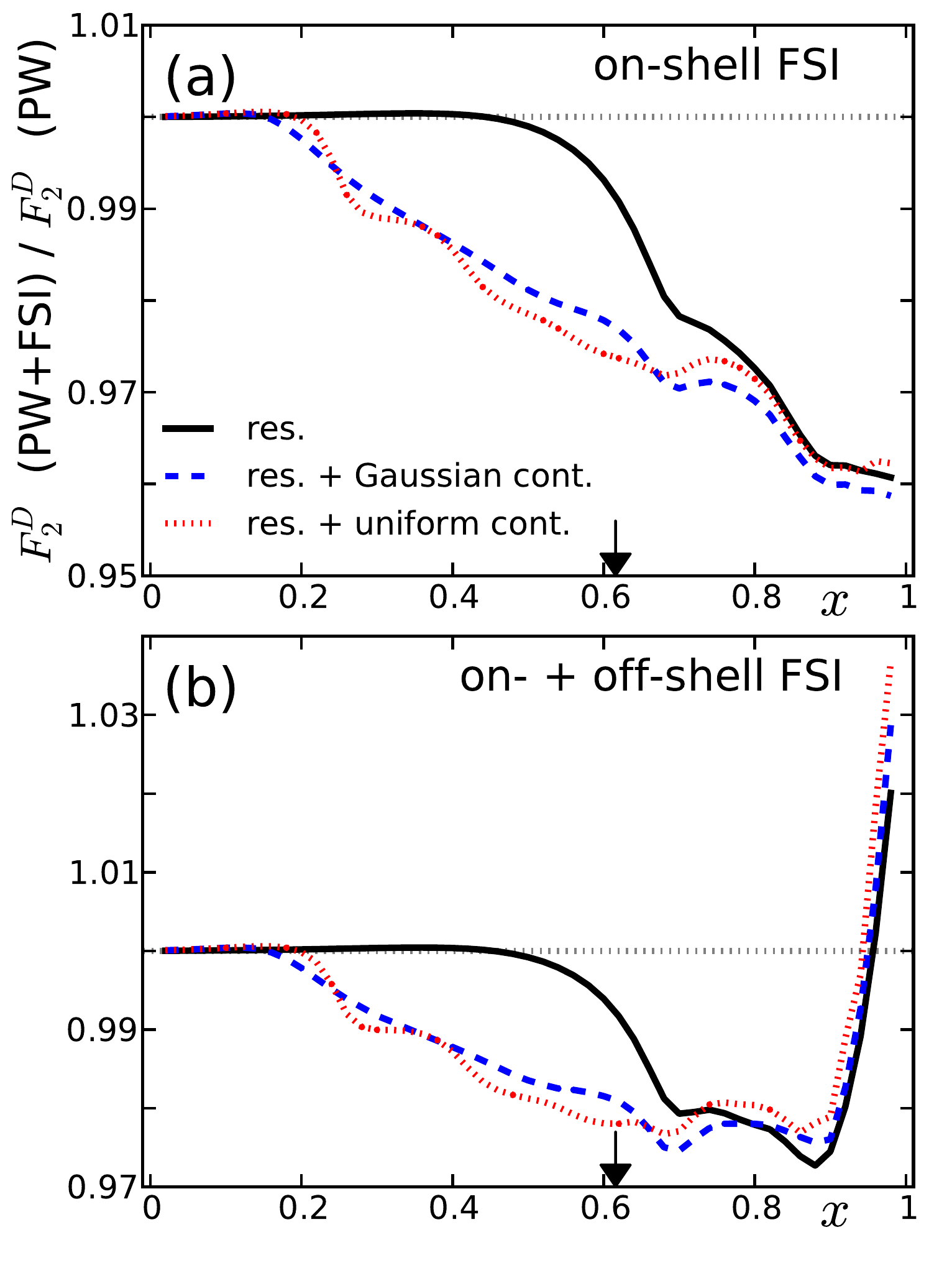}
\caption{(Color online)
	Ratio of the deuteron $F_2^D$ structure function with FSIs
	to that computed in the plane-wave (PW) approximation at
	$Q^2=5~\text{GeV}^2$, using only the on-shell amplitude	in
	Eq.~(\ref{eq:fsifinalon}) {\bf [(a)]} and including
	also the off-shell contribution of Eq.~(\ref{eq:fsifinaloff})
	{\bf [(b)]}.
	The results with the resonance region contributions alone
	(black solid lines) are compared with those including a
	continuum component with a Gaussian distribution (blue dashed)
	and a uniform distribution (red dotted).
	The arrow along the $x$-axis indicates the boundary at
	$W=2$~GeV between the resonance and DIS regions for free
	nucleon kinematics.}
\label{fig:f2fsi_on_off}
\end{center}
\end{figure}

The ratio of the $F_2^D$ structure function evaluated including the
effects of FSIs to that with the plane-wave contribution only is
illustrated in Fig.~\ref{fig:f2fsi_on_off} at $Q^2=5~\text{GeV}^2$.
The results with the three resonance contributions alone are
compared with those that account also for the DIS continuum.
For the shape of the continuum spectral function $\rho(m_{X_2})$ in
Eq.~(\ref{eq:sumres}) we consider two different parametrizations,
based on Gaussian and uniform distributions.
The Gaussian distributions are chosen to be centered at
$m_{X_2} = \{2,\, 2.5,\, 3.4\}$~GeV, with corresponding widths of
$\{ 300,\, 500,\, 700 \}$~MeV, in order to correspond to the values
of the two highest $W$ bins in Ref.~\cite{Klimenko:2005zz}.
The uniform parametrization uses three distributions limited
by $m_{X_2}$ values of $\{ 1.85,\, 2.2,\, 2.8,\, 4 \}$~GeV.
While the two $\rho(m_{X_2})$ models exhibit quite different
shapes, the difference between them has only a very modest effect
on the $F_2^D$ ratios in Fig.~(\ref{fig:f2fsi_on_off}).

The inclusion of the FSI contribution generally suppresses the $F_2^D$
structure function by several percent, particularly at high values of
$x$, where the intermediate states have greatest phase space available.
With the resonance contributions only, and for the on-shell rescattering
amplitude, the structure function is reduced by $\approx 1\%-2\%$ for
$x \approx 0.6-0.7$, and by up to $\approx 3\%-4\%$ for $x \gtrsim 0.8$.
The effect of the continuum contribution is to reduce $F_2^D$ at lower
values of $x$, for masses below the free nucleon threshold of $W=2$~GeV
(corresponding to $x \lesssim 0.6$ at $Q^2=5~\text{GeV}^2$).
The addition of the off-shell FSI scattering amplitude in
Eq.~(\ref{eq:off-ampl}) curtails some of the $F_2^D$ suppression,
to $\approx 2\%$ up to $x \approx 0.9$, with a strong enhancement
of the ratio above unity for $x \to 1$ which is largely independent
of the details of the distribution of the intermediate state masses.

\begin{figure}[t]
\begin{center}
\includegraphics[width=\textwidth]{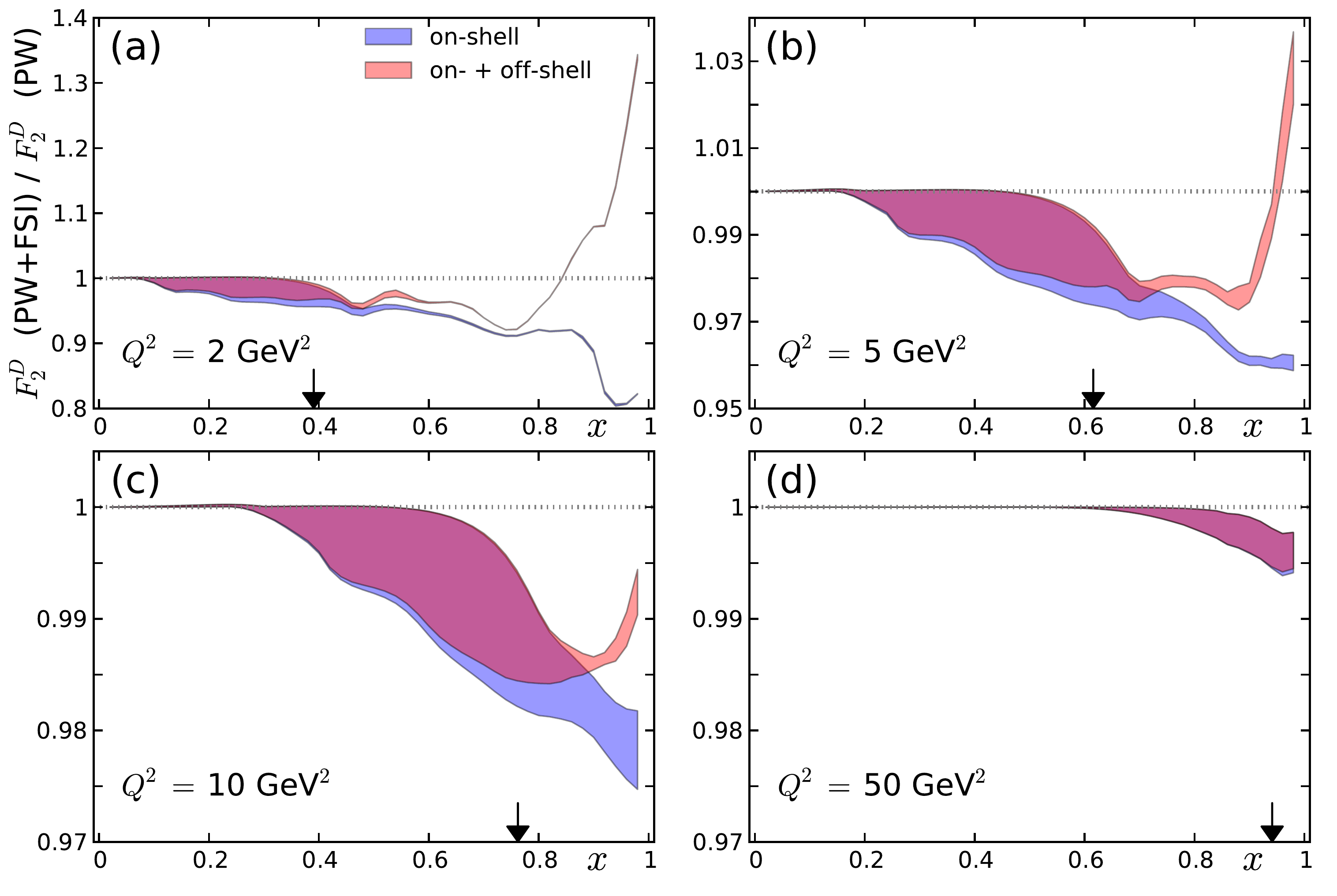}
\caption{(Color online)
	Ratio of the deuteron $F_2^D$ structure function including
	FSIs to that computed in the PW approximation, at several
	values of $Q^2$ from 2~GeV$^2$ to 50~GeV$^2$.
	The on-shell only results from Eq.~(\ref{eq:fsifinalon})
	(blue shaded bands) and those including off-shell contributions
	from Eq.~(\ref{eq:fsifinaloff}) (orange shaded bands) span the
	range of models for the distribution of intermediate state
	masses shown in Fig.~\ref{fig:f2fsi_on_off}.
	The arrows along the $x$-axis indicate the $W=2$~GeV point
	at each $Q^2$ value for free nucleon kinematics.}
\label{fig:f2fsi_Q2}
\end{center}
\end{figure}

The $Q^2$ dependence of the FSI contributions is illustrated in
Fig.~\ref{fig:f2fsi_Q2} for $F_2^D$ calculated in terms of on-shell
only amplitudes and including the off-shell corrections.  Here the
bands envelope the range of intermediate state mass distributions
considered in Fig.~\ref{fig:f2fsi_on_off}, including the resonance
components and the models for the DIS continuum.
For the on-shell part of the FSI amplitude in Eq.~(\ref{eq:fsifinalon})
the largest effect is seen at the lowest $Q^2$ values, where the
invariant masses of the resonances contributing to the FSI amplitude
are attained for $x \approx 0.4-1$, and the $Q^2$ suppression of the
rescattering amplitude is weakest.
The $F^2_D$ structure function ratio in this case is below unity for
all values of $x$, with the largest effect seen at $x \gtrsim 0.9$,
where $F_2^D$ is reduced by up to 20\% at $Q^2=2~\text{GeV}^2$.
At larger $Q^2$, the effect is significantly smaller, with $F_2^D$
suppressed by less than 0.5\% at $Q^2=50~\text{GeV}^2$.

\begin{figure}[t]
\begin{center}
\includegraphics[width=0.55\textwidth]{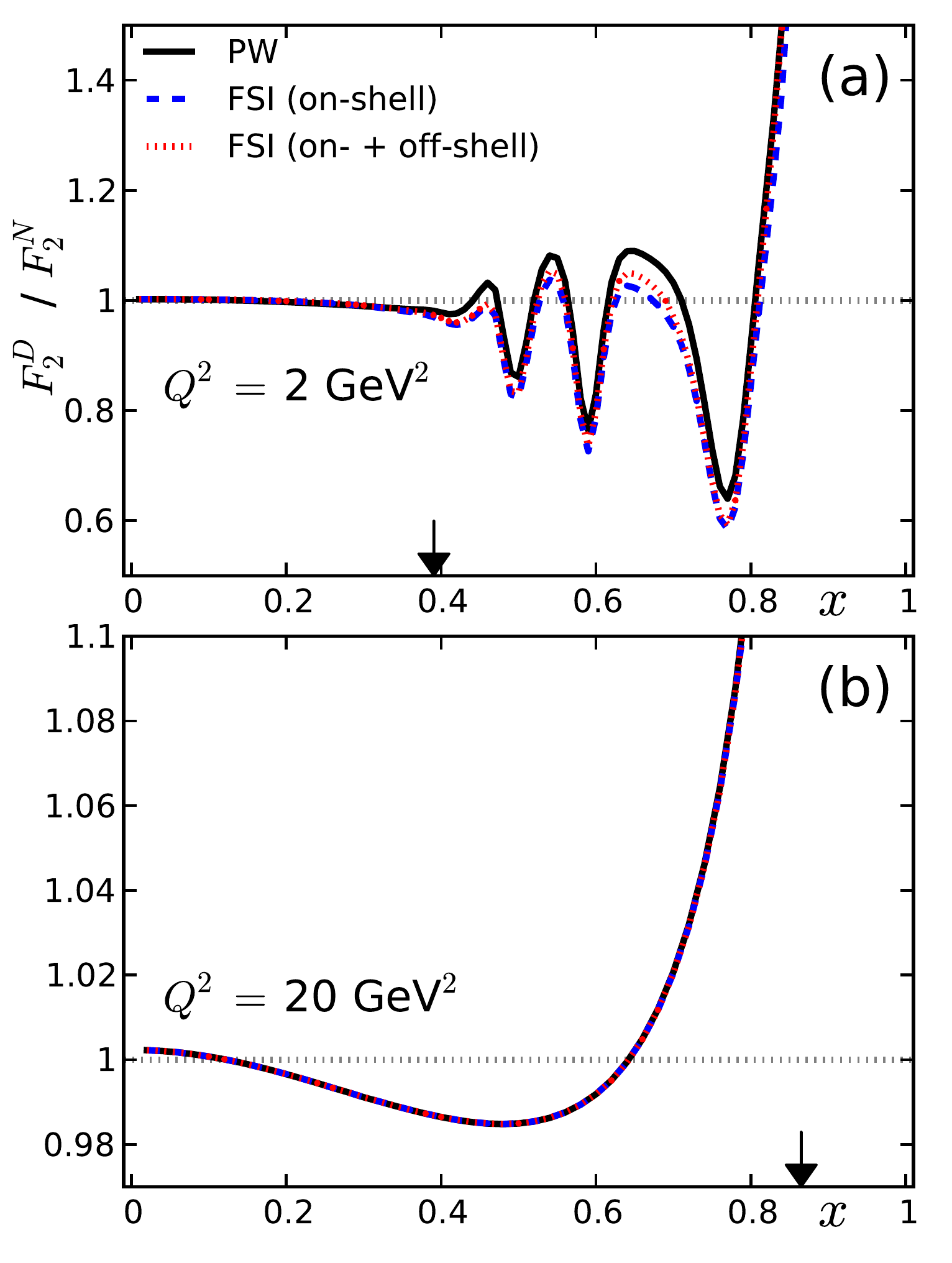}
\caption{(Color online)
	Ratio of the deuteron $F_2^D$ to isoscalar nucleon $F_2^N$
	structure functions at $Q^2 = 2$~GeV$^2$ {\bf [(a)]}
	and $Q^2 = 20$~GeV$^2$ {\bf [(b)]}.
	The PW results (black solid curves) are compared with
	those including FSIs with the on-shell amplitude
	[Eq.~(\ref{eq:fsifinalon})] (blue dashed curves) and
	with the addition of the off-shell contribution of
	[Eq.~(\ref{eq:fsifinaloff})] (red dotted curves).
	The arrows along the $x$-axis indicate the $W=2$~GeV
	boundary for free nucleon kinematics.
	The nucleon resonance structures are clearly visible
	in the ratio at $Q^2=2$~GeV$^2$.}
\label{fig:fDtofN}
\end{center}
\end{figure}

The contribution of the off-shell amplitude has the opposite effect
on $F_2^D$ to that of the on-shell amplitude.  The size of the
off-shell contribution is negligible at low $x$, but becomes
important at high $x$ and low $Q^2$.
This behavior can be expected if one considers a dynamical picture
based on the analysis of the propagation of the virtual particle
in the intermediate state of the reaction \cite{Frankfurt:1993sp}.
As with the on-shell term, the size of the off-shell contribution
becomes smaller with increasing $Q^2$, reflecting the suppression
of FSI effects at large $Q^2$.  In the high-$x$ region all possible
$m_{X_2}$ contributions included in Eq.~(\ref{eq:sumres}) are
kinematically accessible for the phase space of the produced $X$,
so that all possible rescattering contributions can play a role.
We should stress, however, that (as noted above) the curves in
Fig.~\ref{fig:f2fsi_Q2} represent the {\it maximum} possible effect
of the FSI contribution in our model.

\begin{figure}[t]
\begin{center}
\includegraphics[width=0.55\textwidth]{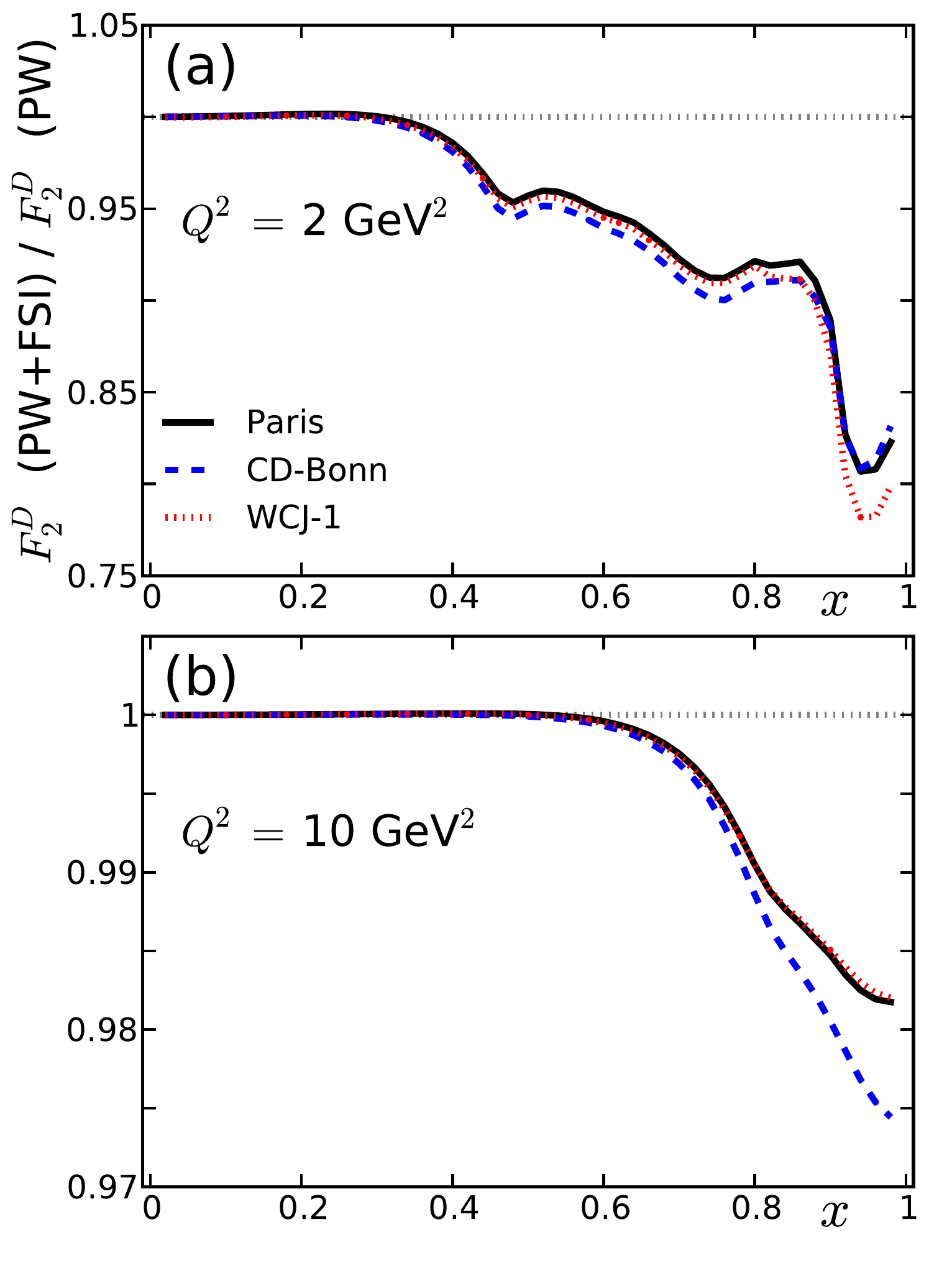}
\caption{(Color online)
	Ratio of the deuteron $F_2^D$ structure function with
	and without FSIs at $Q^2 = 2$~GeV$^2$ {\bf [(a)]}
        and $Q^2 = 20$~GeV$^2$ {\bf [(b)]}.
	The results using the Paris deuteron wave function
	\cite{Lacombe:1980dr} (black solid curves) are compared
	with those employing the CD-Bonn \cite{Machleidt:2000ge}
	(blue dashed curves) and WCJ-1 \cite{Gross:2008ps}
	(red dotted curves) wave functions.}
\label{fig:deutwf}
\end{center}
\end{figure}

To put the effects of the FSIs in a more familiar context, in
Fig.~\ref{fig:fDtofN} we show the ratio of the deuteron $F_2^D$ to
isoscalar nucleon $F_2^N$ structure functions at $Q^2 = 2$~GeV$^2$
and 20~GeV$^2$, for the PW calculation and with FSIs using the
on-shell and off-shell amplitudes.
At the lower $Q^2$ value the resonance structures are clearly
visible at large $x$ ($x \gtrsim 0.4$), with the ratios including
FSIs slightly lower in magnitude compared with the PW results.
With increasing $Q^2$ the resonance peaks migrate to higher $x$,
revealing a smooth ratio with a trough at moderate $x$ values
($x \sim 0.6$) and a large enhancement due to Fermi motion at
$x \to 1$, typical of the ``nuclear EMC effect'' \cite{Melnitchouk:1993nk,
Kulagin:1994fz, Sargsian:2001gu, Kulagin:2004ie, CJ11}.
By $Q^2=20~\text{GeV}^2$ the impact of the FSIs on the structure
function ratio in the DIS region is negligible.

The sensitivity of these results to the specific model of the
deuteron wave function used in the calculation can be seen in
Fig.~\ref{fig:deutwf}, where the ratios of the $F_2^D$ structure
functions with and without FSIs are shown for the Paris
\cite{Lacombe:1980dr}, CD-Bonn \cite{Machleidt:2000ge},
and WCJ-1 \cite{Gross:2008ps} $NN$ potential models.
For comparison, we consider here the rescattering model
involving only FSI contributions from the resonance region.
This choice of wave functions spans the maximal range of
effects possible using modern, high-precision $NN$ potentials
\cite{CJ11}.
Except at very high $x$ values ($x \gtrsim 0.8$), where the
$F_2^D$ is most sensitive to the short-range structure of the
deuteron, the effects of the different deuteron wave functions
is very small, in both the resonance and DIS kinematics.
This result gives us some confidence that the findings of the
FSI calculation are relatively robust, and not strongly dependent
on the details of the deuteron structure.

%%%%%%%%%%%%%%%%%%%%%%%%%%%%%%%%%%%%%%%%%%%%%%%%%%%%%%%%%%%%%%%%%%%%%%%%%
\section{Conclusion} \label{sec:conclusion}

In this work we have presented a first quantitative analysis of the
effects of FSIs in inclusive DIS from deuterium.  Using the optical
theorem and the properties of high-energy diffractive rescattering,
a general result was derived within the generalized eikonal
approximation for the FSI contribution to the inclusive DIS deuteron
cross section.  To obtain numerical estimates of the FSI corrections
we introduced a factorized model expressed through the bound-nucleon
inelastic structure functions and the sum of hadronic rescattering
amplitudes.  The contributions to the rescattering amplitudes were
modeled in terms of a set of three resonances with masses $W<2$~GeV
and a continuum spectrum for the DIS region at $W \geq 2$~GeV.
The particular structure of these intermediate states was motivated
by recent analyses \cite{Cosyn:2010ux} of spectator proton production
data in SIDIS from the deuteron \cite{Klimenko:2005zz}.

The formalism developed here includes on-shell and off-shell
contributions to the rescattering amplitude, the latter which
introduces some degree of model dependence.
Importantly, however, we find that within this framework the FSI
corrections vanish in the limit of large $Q^2$ due to phase space
constraints, independent of the details of the intermediate
hadronic states.
Numerically, we find sizeable FSI contributions to the inclusive
$F_2^D$ structure function at $Q^2 \lesssim 10~\text{GeV}^2$ and
$x \gtrsim 0.4$.  Generally, the on-shell rescattering amplitude
lowers the value of the structure function, with the effects most
prominent at high $x$ and low $Q^2$, with a decrease of up to 20\%
found at $Q^2=2~\text{GeV}^2$ as $x \to 1$.

The off-shell contributions to the rescattering generally enter
with the opposite sign relative to the on-shell contributions,
and in the $x \gtrsim 0.8$ region where their effects are most
important they increase the magnitude of the deuteron structure
function.  Estimating the parameters that determine the off-shell
contribution is difficult, however, and in our calculations we use
the difference between the on-shell and off-shell results as an
estimate on the uncertainty of the FSI corrections.  In contrast,
the dependence of the results on the deuteron wave function is found
to be significantly smaller than the typical size of the FSI effects.

Our overall conclusion is that at $x \gtrsim 0.6$ and
$Q^2 \lesssim 10~\text{GeV}^2$ the FSI effects can contribute to
the deuteron $F_2^D$ structure function at the level of $2\%-5$\%,
and should be considered in extractions of the neutron structure
function from inclusive deuteron data at low $Q^2$.
At larger $Q^2$ values ($Q^2 \gtrsim 10~\text{GeV}^2$) in the
deep-inelastic region the FSI effects are found to be negligible.
Future data on tagged structure functions in SIDIS from the deuteron
over a broader range of $Q^2$ and $W$ would be very helpful in
further constraining the shape and magnitude of the FSI contributions
to inclusive deuteron DIS.  In this respect the planned ``BONuS''
experiment \cite{BONUS12} at the energy-upgraded CEBAF accelerator
at Jefferson Lab has the potential to provide new empirical guidance.

%%%%%%%%%%%%%%%%%%%%%%%%%%%%%%%%%%%%%%%%%%%%%%%%%%%%%%%%%%%%%%%%%%%%%%%%%
\subsection*{Acknowledgments}

W.C. is supported by the Research Foundation Flanders (FWO-Flanders).
W.M. is supported by the DOE contract No. DE-AC05-06OR23177,
under which Jefferson Science Associates, LLC operates Jefferson Lab.
M.S. is supported by U.S. Department of Energy grant under contract
DE-FG02-01ER41172.
The computational resources (Stevin Supercomputer Infrastructure)
and services used in this work were provided by Ghent University,
the Hercules Foundation and the Flemish Government -- department EWI.

\newpage
%%%%%%%%%%%%%%%%%%%%%%%%%%%%%%%%%%%%%%%%%%%%%%%%%%%%%%%%%%%%%%%%%%%%%%%%%
\bibliography{inclref.bib}

\begin{thebibliography}{61}
\expandafter\ifx\csname natexlab\endcsname\relax\def\natexlab#1{#1}\fi
\expandafter\ifx\csname bibnamefont\endcsname\relax
  \def\bibnamefont#1{#1}\fi
\expandafter\ifx\csname bibfnamefont\endcsname\relax
  \def\bibfnamefont#1{#1}\fi
\expandafter\ifx\csname citenamefont\endcsname\relax
  \def\citenamefont#1{#1}\fi
\expandafter\ifx\csname url\endcsname\relax
  \def\url#1{\texttt{#1}}\fi
\expandafter\ifx\csname urlprefix\endcsname\relax\def\urlprefix{URL }\fi
\providecommand{\bibinfo}[2]{#2}
\providecommand{\eprint}[2][]{\url{#2}}

\bibitem[{\citenamefont{Whitlow et~al.}(1992)\citenamefont{Whitlow, Riordan,
  Dasu, Rock, and Bodek}}]{Whitlow:1991uw}
\bibinfo{author}{\bibfnamefont{L.~W.} \bibnamefont{Whitlow}},
  \bibinfo{author}{\bibfnamefont{E.~M.} \bibnamefont{Riordan}},
  \bibinfo{author}{\bibfnamefont{S.}~\bibnamefont{Dasu}},
  \bibinfo{author}{\bibfnamefont{S.}~\bibnamefont{Rock}}, \bibnamefont{and}
  \bibinfo{author}{\bibfnamefont{A.}~\bibnamefont{Bodek}},
  \bibinfo{journal}{Phys. Lett. B} \textbf{\bibinfo{volume}{282}},
  \bibinfo{pages}{475} (\bibinfo{year}{1992}).

\bibitem[{\citenamefont{Benvenuti et~al.}(1990)}]{Benvenuti1990592}
\bibinfo{author}{\bibfnamefont{A.~C.} \bibnamefont{Benvenuti}}
  \bibnamefont{et~al.}, \bibinfo{journal}{Phys. Lett. B}
  \textbf{\bibinfo{volume}{237}}, \bibinfo{pages}{592} (\bibinfo{year}{1990}).

\bibitem[{\citenamefont{Arneodo et~al.}(1997{\natexlab{a}})}]{Arneodo:1996kd}
\bibinfo{author}{\bibfnamefont{M.}~\bibnamefont{Arneodo}} \bibnamefont{et~al.},
  \bibinfo{journal}{Nucl. Phys.} \textbf{\bibinfo{volume}{B487}},
  \bibinfo{pages}{3} (\bibinfo{year}{1997}{\natexlab{a}}).

\bibitem[{\citenamefont{Arneodo et~al.}(1997{\natexlab{b}})}]{Arneodo:1996qe}
\bibinfo{author}{\bibfnamefont{M.}~\bibnamefont{Arneodo}} \bibnamefont{et~al.},
  \bibinfo{journal}{Nucl. Phys.} \textbf{\bibinfo{volume}{B483}},
  \bibinfo{pages}{3} (\bibinfo{year}{1997}{\natexlab{b}}).

\bibitem[{\citenamefont{Tvaskis et~al.}(2010)}]{Tvaskis:2010as}
\bibinfo{author}{\bibfnamefont{V.}~\bibnamefont{Tvaskis}} \bibnamefont{et~al.},
  \bibinfo{journal}{Phys. Rev. C} \textbf{\bibinfo{volume}{81}},
  \bibinfo{pages}{055207} (\bibinfo{year}{2010}).

\bibitem[{\citenamefont{Accardi et~al.}(2010)}]{CJ10}
\bibinfo{author}{\bibfnamefont{A.}~\bibnamefont{Accardi}} \bibnamefont{et~al.},
  \bibinfo{journal}{Phys. Rev. D} \textbf{\bibinfo{volume}{81}},
  \bibinfo{pages}{034016} (\bibinfo{year}{2010}).

\bibitem[{\citenamefont{Accardi et~al.}(2011)}]{CJ11}
\bibinfo{author}{\bibfnamefont{A.}~\bibnamefont{Accardi}} \bibnamefont{et~al.},
  \bibinfo{journal}{Phys. Rev. D} \textbf{\bibinfo{volume}{84}},
  \bibinfo{pages}{014008} (\bibinfo{year}{2011}).

\bibitem[{\citenamefont{Owens et~al.}(2013)\citenamefont{Owens, Accardi, and
  Melnitchouk}}]{CJ12}
\bibinfo{author}{\bibfnamefont{J.~F.} \bibnamefont{Owens}},
  \bibinfo{author}{\bibfnamefont{A.}~\bibnamefont{Accardi}}, \bibnamefont{and}
  \bibinfo{author}{\bibfnamefont{W.}~\bibnamefont{Melnitchouk}},
  \bibinfo{journal}{Phys. Rev. D} \textbf{\bibinfo{volume}{87}},
  \bibinfo{pages}{094012} (\bibinfo{year}{2013}).

\bibitem[{\citenamefont{Malace et~al.}(2010)\citenamefont{Malace, Kahn,
  Melnitchouk, and Keppel}}]{Malace:2009dg}
\bibinfo{author}{\bibfnamefont{S.~P.} \bibnamefont{Malace}},
  \bibinfo{author}{\bibfnamefont{Y.}~\bibnamefont{Kahn}},
  \bibinfo{author}{\bibfnamefont{W.}~\bibnamefont{Melnitchouk}},
  \bibnamefont{and} \bibinfo{author}{\bibfnamefont{C.~E.}
  \bibnamefont{Keppel}}, \bibinfo{journal}{Phys. Rev. Lett.}
  \textbf{\bibinfo{volume}{104}}, \bibinfo{pages}{102001}
  (\bibinfo{year}{2010}).

\bibitem[{\citenamefont{Martin et~al.}(2013)}]{Martin:2012da}
\bibinfo{author}{\bibfnamefont{A.~D.} \bibnamefont{Martin}}
  \bibnamefont{et~al.}, \bibinfo{journal}{Eur. Phys. J. C}
  \textbf{\bibinfo{volume}{73}}, \bibinfo{pages}{2318} (\bibinfo{year}{2013}).

\bibitem[{\citenamefont{Ball et~al.}(2013)}]{Ball:2013gsa}
\bibinfo{author}{\bibfnamefont{R.~D.} \bibnamefont{Ball}} \bibnamefont{et~al.},
  \bibinfo{journal}{Phys. Lett. B} \textbf{\bibinfo{volume}{723}},
  \bibinfo{pages}{330} (\bibinfo{year}{2013}).

\bibitem[{\citenamefont{Frankfurt and Strikman}(1978)}]{Frankfurt:1976hb}
\bibinfo{author}{\bibfnamefont{L.}~\bibnamefont{Frankfurt}} \bibnamefont{and}
  \bibinfo{author}{\bibfnamefont{M.}~\bibnamefont{Strikman}},
  \bibinfo{journal}{Phys. Lett. B} \textbf{\bibinfo{volume}{76}},
  \bibinfo{pages}{333} (\bibinfo{year}{1978}).

\bibitem[{\citenamefont{Kulagin et~al.}(1994)\citenamefont{Kulagin, Piller, and
  Weise}}]{Kulagin:1994fz}
\bibinfo{author}{\bibfnamefont{S.~A.} \bibnamefont{Kulagin}},
  \bibinfo{author}{\bibfnamefont{G.}~\bibnamefont{Piller}}, \bibnamefont{and}
  \bibinfo{author}{\bibfnamefont{W.}~\bibnamefont{Weise}},
  \bibinfo{journal}{Phys. Rev. C} \textbf{\bibinfo{volume}{50}},
  \bibinfo{pages}{1154} (\bibinfo{year}{1994}).

\bibitem[{\citenamefont{Melnitchouk and Thomas}(1996)}]{Melnitchouk:1995fc}
\bibinfo{author}{\bibfnamefont{W.}~\bibnamefont{Melnitchouk}} \bibnamefont{and}
  \bibinfo{author}{\bibfnamefont{A.~W.} \bibnamefont{Thomas}},
  \bibinfo{journal}{Phys. Lett. B} \textbf{\bibinfo{volume}{377}},
  \bibinfo{pages}{11} (\bibinfo{year}{1996}).

\bibitem[{\citenamefont{Afnan et~al.}(2000)}]{Afnan:2000uh}
\bibinfo{author}{\bibfnamefont{I.~R.} \bibnamefont{Afnan}}
  \bibnamefont{et~al.}, \bibinfo{journal}{Phys. Lett. B}
  \textbf{\bibinfo{volume}{493}}, \bibinfo{pages}{36} (\bibinfo{year}{2000}).

\bibitem[{\citenamefont{Afnan et~al.}(2003)}]{Afnan:2003vh}
\bibinfo{author}{\bibfnamefont{I.~R.} \bibnamefont{Afnan}}
  \bibnamefont{et~al.}, \bibinfo{journal}{Phys. Rev. C}
  \textbf{\bibinfo{volume}{68}}, \bibinfo{pages}{035201}
  (\bibinfo{year}{2003}).

\bibitem[{\citenamefont{Sargsian et~al.}(2002)\citenamefont{Sargsian, Simula,
  and Strikman}}]{Sargsian:2001gu}
\bibinfo{author}{\bibfnamefont{M.}~\bibnamefont{Sargsian}},
  \bibinfo{author}{\bibfnamefont{S.}~\bibnamefont{Simula}}, \bibnamefont{and}
  \bibinfo{author}{\bibfnamefont{M.}~\bibnamefont{Strikman}},
  \bibinfo{journal}{Phys. Rev. C} \textbf{\bibinfo{volume}{66}},
  \bibinfo{pages}{024001} (\bibinfo{year}{2002}).

\bibitem[{\citenamefont{Kulagin and Petti}(2006)}]{Kulagin:2004ie}
\bibinfo{author}{\bibfnamefont{S.~A.} \bibnamefont{Kulagin}} \bibnamefont{and}
  \bibinfo{author}{\bibfnamefont{R.}~\bibnamefont{Petti}},
  \bibinfo{journal}{Nucl. Phys.} \textbf{\bibinfo{volume}{A765}},
  \bibinfo{pages}{126} (\bibinfo{year}{2006}).

\bibitem[{\citenamefont{Kahn et~al.}(2009)\citenamefont{Kahn, Melnitchouk, and
  Kulagin}}]{Kahn:2008nq}
\bibinfo{author}{\bibfnamefont{Y.}~\bibnamefont{Kahn}},
  \bibinfo{author}{\bibfnamefont{W.}~\bibnamefont{Melnitchouk}},
  \bibnamefont{and} \bibinfo{author}{\bibfnamefont{S.~A.}
  \bibnamefont{Kulagin}}, \bibinfo{journal}{Phys. Rev. C}
  \textbf{\bibinfo{volume}{79}}, \bibinfo{pages}{035205}
  (\bibinfo{year}{2009}).

\bibitem[{\citenamefont{Arrington et~al.}(2012)\citenamefont{Arrington, Rubin,
  and Melnitchouk}}]{Arrington:2011qt}
\bibinfo{author}{\bibfnamefont{J.}~\bibnamefont{Arrington}},
  \bibinfo{author}{\bibfnamefont{J.~G.} \bibnamefont{Rubin}}, \bibnamefont{and}
  \bibinfo{author}{\bibfnamefont{W.}~\bibnamefont{Melnitchouk}},
  \bibinfo{journal}{Phys. Rev. Lett.} \textbf{\bibinfo{volume}{108}},
  \bibinfo{pages}{252001} (\bibinfo{year}{2012}).

\bibitem[{\citenamefont{Frankfurt and Strikman}(1989)}]{Frankfurt:1988zg}
\bibinfo{author}{\bibfnamefont{L.}~\bibnamefont{Frankfurt}} \bibnamefont{and}
  \bibinfo{author}{\bibfnamefont{M.}~\bibnamefont{Strikman}},
  \bibinfo{journal}{Nucl. Phys.} \textbf{\bibinfo{volume}{B316}},
  \bibinfo{pages}{340} (\bibinfo{year}{1989}).

\bibitem[{\citenamefont{Nikolaev and Zakharov}(1991)}]{Nikolaev:1990ja}
\bibinfo{author}{\bibfnamefont{N.~N.} \bibnamefont{Nikolaev}} \bibnamefont{and}
  \bibinfo{author}{\bibfnamefont{B.~G.} \bibnamefont{Zakharov}},
  \bibinfo{journal}{Z. Phys. C} \textbf{\bibinfo{volume}{49}},
  \bibinfo{pages}{607} (\bibinfo{year}{1991}).

\bibitem[{\citenamefont{Zoller}(1992)}]{Zoller:1991ph}
\bibinfo{author}{\bibfnamefont{V.~R.} \bibnamefont{Zoller}},
  \bibinfo{journal}{Z. Phys. C} \textbf{\bibinfo{volume}{54}},
  \bibinfo{pages}{425} (\bibinfo{year}{1992}).

\bibitem[{\citenamefont{Badelek and Kwiecinski}(1992)}]{Badelek:1991qa}
\bibinfo{author}{\bibfnamefont{B.}~\bibnamefont{Badelek}} \bibnamefont{and}
  \bibinfo{author}{\bibfnamefont{J.}~\bibnamefont{Kwiecinski}},
  \bibinfo{journal}{Nucl. Phys.} \textbf{\bibinfo{volume}{B370}},
  \bibinfo{pages}{278} (\bibinfo{year}{1992}).

\bibitem[{\citenamefont{Melnitchouk and Thomas}(1993)}]{Melnitchouk:1992eu}
\bibinfo{author}{\bibfnamefont{W.}~\bibnamefont{Melnitchouk}} \bibnamefont{and}
  \bibinfo{author}{\bibfnamefont{A.~W.} \bibnamefont{Thomas}},
  \bibinfo{journal}{Phys. Rev. D} \textbf{\bibinfo{volume}{47}},
  \bibinfo{pages}{3783} (\bibinfo{year}{1993}).

\bibitem[{\citenamefont{Piller and Weise}(2000)}]{Piller:1999wx}
\bibinfo{author}{\bibfnamefont{G.}~\bibnamefont{Piller}} \bibnamefont{and}
  \bibinfo{author}{\bibfnamefont{W.}~\bibnamefont{Weise}},
  \bibinfo{journal}{Phys. Rep.} \textbf{\bibinfo{volume}{330}},
  \bibinfo{pages}{1} (\bibinfo{year}{2000}).

\bibitem[{\citenamefont{Melnitchouk et~al.}(2005)\citenamefont{Melnitchouk,
  Ent, and Keppel}}]{Melnitchouk:2005zr}
\bibinfo{author}{\bibfnamefont{W.}~\bibnamefont{Melnitchouk}},
  \bibinfo{author}{\bibfnamefont{R.}~\bibnamefont{Ent}}, \bibnamefont{and}
  \bibinfo{author}{\bibfnamefont{C.~E.} \bibnamefont{Keppel}},
  \bibinfo{journal}{Phys. Rep.} \textbf{\bibinfo{volume}{406}},
  \bibinfo{pages}{127} (\bibinfo{year}{2005}).

\bibitem[{\citenamefont{Collins}(2011)}]{Collins:2011zzd}
\bibinfo{author}{\bibfnamefont{J.~C.} \bibnamefont{Collins}},
  \emph{\bibinfo{title}{Foundations of perturbative QCD}}, Cambridge monographs
  on particle physics, nuclear physics, and cosmology
  (\bibinfo{publisher}{Cambridge Univ. Press}, \bibinfo{address}{New York, NY},
  \bibinfo{year}{2011}).

\bibitem[{\citenamefont{Ciofi~degli Atti et~al.}(2004)\citenamefont{Ciofi~degli
  Atti, Kaptari, and Kopeliovich}}]{Ciofi:2003pb}
\bibinfo{author}{\bibfnamefont{C.}~\bibnamefont{Ciofi~degli Atti}},
  \bibinfo{author}{\bibfnamefont{L.~P.} \bibnamefont{Kaptari}},
  \bibnamefont{and} \bibinfo{author}{\bibfnamefont{B.~Z.}
  \bibnamefont{Kopeliovich}}, \bibinfo{journal}{Eur. Phys. J. A}
  \textbf{\bibinfo{volume}{19}}, \bibinfo{pages}{145} (\bibinfo{year}{2004}).

\bibitem[{\citenamefont{Ciofi~degli Atti et~al.}(2001)\citenamefont{Ciofi~degli
  Atti, Kaptari, and Treleani}}]{Ciofi:2000xj}
\bibinfo{author}{\bibfnamefont{C.}~\bibnamefont{Ciofi~degli Atti}},
  \bibinfo{author}{\bibfnamefont{L.~P.} \bibnamefont{Kaptari}},
  \bibnamefont{and} \bibinfo{author}{\bibfnamefont{D.}~\bibnamefont{Treleani}},
  \bibinfo{journal}{Phys. Rev. C} \textbf{\bibinfo{volume}{63}},
  \bibinfo{pages}{044601} (\bibinfo{year}{2001}).

\bibitem[{\citenamefont{Palli et~al.}(2009)\citenamefont{Palli, Ciofi~degli
  Atti, Kaptari, Mezzetti, and Alvioli}}]{Palli:2009it}
\bibinfo{author}{\bibfnamefont{V.}~\bibnamefont{Palli}},
  \bibinfo{author}{\bibfnamefont{C.}~\bibnamefont{Ciofi~degli Atti}},
  \bibinfo{author}{\bibfnamefont{L.~P.} \bibnamefont{Kaptari}},
  \bibinfo{author}{\bibfnamefont{C.~B.} \bibnamefont{Mezzetti}},
  \bibnamefont{and} \bibinfo{author}{\bibfnamefont{M.}~\bibnamefont{Alvioli}},
  \bibinfo{journal}{Phys. Rev. C} \textbf{\bibinfo{volume}{80}},
  \bibinfo{pages}{054610} (\bibinfo{year}{2009}).

\bibitem[{\citenamefont{Cosyn and Sargsian}(2011{\natexlab{a}})}]{Cosyn:2010ux}
\bibinfo{author}{\bibfnamefont{W.}~\bibnamefont{Cosyn}} \bibnamefont{and}
  \bibinfo{author}{\bibfnamefont{M.}~\bibnamefont{Sargsian}},
  \bibinfo{journal}{Phys. Rev. C} \textbf{\bibinfo{volume}{84}},
  \bibinfo{pages}{014601} (\bibinfo{year}{2011}{\natexlab{a}}).

\bibitem[{\citenamefont{Cosyn and Sargsian}(2011{\natexlab{b}})}]{Cosyn:2011jc}
\bibinfo{author}{\bibfnamefont{W.}~\bibnamefont{Cosyn}} \bibnamefont{and}
  \bibinfo{author}{\bibfnamefont{M.}~\bibnamefont{Sargsian}},
  \bibinfo{journal}{AIP Conf. Proc.} \textbf{\bibinfo{volume}{1369}},
  \bibinfo{pages}{121} (\bibinfo{year}{2011}{\natexlab{b}}).

\bibitem[{\citenamefont{Klimenko et~al.}(2006)}]{Klimenko:2005zz}
\bibinfo{author}{\bibfnamefont{A.~V.} \bibnamefont{Klimenko}}
  \bibnamefont{et~al.}, \bibinfo{journal}{Phys. Rev. C}
  \textbf{\bibinfo{volume}{73}}, \bibinfo{pages}{035212}
  (\bibinfo{year}{2006}).

\bibitem[{\citenamefont{Baillie et~al.}(2012)}]{Baillie:2011za}
\bibinfo{author}{\bibfnamefont{N.}~\bibnamefont{Baillie}} \bibnamefont{et~al.},
  \bibinfo{journal}{Phys. Rev. Lett.} \textbf{\bibinfo{volume}{108}},
  \bibinfo{pages}{199902} (\bibinfo{year}{2012}).

\bibitem[{\citenamefont{Frankfurt et~al.}(1997)\citenamefont{Frankfurt,
  Sargsian, and Strikman}}]{Frankfurt:1996xx}
\bibinfo{author}{\bibfnamefont{L.}~\bibnamefont{Frankfurt}},
  \bibinfo{author}{\bibfnamefont{M.}~\bibnamefont{Sargsian}}, \bibnamefont{and}
  \bibinfo{author}{\bibfnamefont{M.}~\bibnamefont{Strikman}},
  \bibinfo{journal}{Phys. Rev. C} \textbf{\bibinfo{volume}{56}},
  \bibinfo{pages}{1124} (\bibinfo{year}{1997}).

\bibitem[{\citenamefont{Sargsian}(2001)}]{Sargsian:2001ax}
\bibinfo{author}{\bibfnamefont{M.}~\bibnamefont{Sargsian}},
  \bibinfo{journal}{Int. J. Mod. Phys.} \textbf{\bibinfo{volume}{10}},
  \bibinfo{pages}{405} (\bibinfo{year}{2001}).

\bibitem[{\citenamefont{Abramovsky et~al.}(1973)\citenamefont{Abramovsky,
  Gribov, and Kancheli}}]{Abramovsky:1973fm}
\bibinfo{author}{\bibfnamefont{V.~A.} \bibnamefont{Abramovsky}},
  \bibinfo{author}{\bibfnamefont{V.~N.} \bibnamefont{Gribov}},
  \bibnamefont{and} \bibinfo{author}{\bibfnamefont{O.~V.}
  \bibnamefont{Kancheli}}, \bibinfo{journal}{Yad. Fiz.}
  \textbf{\bibinfo{volume}{18}}, \bibinfo{pages}{595} (\bibinfo{year}{1973}).

\bibitem[{\citenamefont{Frankfurt et~al.}(2008)\citenamefont{Frankfurt,
  Sargsian, and Strikman}}]{Frankfurt:2008zv}
\bibinfo{author}{\bibfnamefont{L.}~\bibnamefont{Frankfurt}},
  \bibinfo{author}{\bibfnamefont{M.}~\bibnamefont{Sargsian}}, \bibnamefont{and}
  \bibinfo{author}{\bibfnamefont{M.}~\bibnamefont{Strikman}},
  \bibinfo{journal}{Int. J. Mod. Phys. A} \textbf{\bibinfo{volume}{23}},
  \bibinfo{pages}{2991} (\bibinfo{year}{2008}).

\bibitem[{\citenamefont{Sargsian}(2010)}]{Sargsian:2009hf}
\bibinfo{author}{\bibfnamefont{M.}~\bibnamefont{Sargsian}},
  \bibinfo{journal}{Phys. Rev. C} \textbf{\bibinfo{volume}{82}},
  \bibinfo{pages}{014612} (\bibinfo{year}{2010}).

\bibitem[{\citenamefont{Gross et~al.}(1992)\citenamefont{Gross, Van~Orden, and
  Holinde}}]{Gross:1991pm}
\bibinfo{author}{\bibfnamefont{F.}~\bibnamefont{Gross}},
  \bibinfo{author}{\bibfnamefont{J.~W.} \bibnamefont{Van~Orden}},
  \bibnamefont{and} \bibinfo{author}{\bibfnamefont{K.}~\bibnamefont{Holinde}},
  \bibinfo{journal}{Phys. Rev. C} \textbf{\bibinfo{volume}{45}},
  \bibinfo{pages}{2094} (\bibinfo{year}{1992}).

\bibitem[{\citenamefont{Gribov}(1970)}]{Gribov:1968gs}
\bibinfo{author}{\bibfnamefont{V.~N.} \bibnamefont{Gribov}},
  \bibinfo{journal}{Sov. Phys. JETP} \textbf{\bibinfo{volume}{30}},
  \bibinfo{pages}{709} (\bibinfo{year}{1970}).

\bibitem[{\citenamefont{Bertocchi}(1972)}]{Bertocchi:1972cj}
\bibinfo{author}{\bibfnamefont{L.}~\bibnamefont{Bertocchi}},
  \bibinfo{journal}{Nuovo Cim. A} \textbf{\bibinfo{volume}{11}},
  \bibinfo{pages}{45} (\bibinfo{year}{1972}).

\bibitem[{\citenamefont{Melnitchouk et~al.}(1994)\citenamefont{Melnitchouk,
  Schreiber, and Thomas}}]{Melnitchouk:1993nk}
\bibinfo{author}{\bibfnamefont{W.}~\bibnamefont{Melnitchouk}},
  \bibinfo{author}{\bibfnamefont{A.~W.} \bibnamefont{Schreiber}},
  \bibnamefont{and} \bibinfo{author}{\bibfnamefont{A.~W.}
  \bibnamefont{Thomas}}, \bibinfo{journal}{Phys. Rev. D}
  \textbf{\bibinfo{volume}{49}}, \bibinfo{pages}{1183} (\bibinfo{year}{1994}).

\bibitem[{\citenamefont{Pandharipande and Pieper}(1992)}]{Pandharipande:1992zz}
\bibinfo{author}{\bibfnamefont{V.~R.} \bibnamefont{Pandharipande}}
  \bibnamefont{and} \bibinfo{author}{\bibfnamefont{S.~C.}
  \bibnamefont{Pieper}}, \bibinfo{journal}{Phys. Rev. C}
  \textbf{\bibinfo{volume}{45}}, \bibinfo{pages}{791} (\bibinfo{year}{1992}).

\bibitem[{\citenamefont{Laget}(2005)}]{Laget:2004sm}
\bibinfo{author}{\bibfnamefont{J.-M.} \bibnamefont{Laget}},
  \bibinfo{journal}{Phys. Lett. B} \textbf{\bibinfo{volume}{609}},
  \bibinfo{pages}{49} (\bibinfo{year}{2005}).

\bibitem[{\citenamefont{Ryckebusch et~al.}(2003)\citenamefont{Ryckebusch,
  Debruyne, Lava, Janssen, Van~Overmeire, and
  Van~Cauteren}}]{Ryckebusch:2003fc}
\bibinfo{author}{\bibfnamefont{J.}~\bibnamefont{Ryckebusch}},
  \bibinfo{author}{\bibfnamefont{D.}~\bibnamefont{Debruyne}},
  \bibinfo{author}{\bibfnamefont{P.}~\bibnamefont{Lava}},
  \bibinfo{author}{\bibfnamefont{S.}~\bibnamefont{Janssen}},
  \bibinfo{author}{\bibfnamefont{B.}~\bibnamefont{Van~Overmeire}},
  \bibnamefont{and}
  \bibinfo{author}{\bibfnamefont{T.}~\bibnamefont{Van~Cauteren}},
  \bibinfo{journal}{Nucl. Phys.} \textbf{\bibinfo{volume}{A728}},
  \bibinfo{pages}{226} (\bibinfo{year}{2003}).

\bibitem[{\citenamefont{Aubert et~al.}(1983)}]{Aubert:1983xm}
\bibinfo{author}{\bibfnamefont{J.~J.} \bibnamefont{Aubert}}
  \bibnamefont{et~al.}, \bibinfo{journal}{Phys. Lett. B}
  \textbf{\bibinfo{volume}{123}}, \bibinfo{pages}{275} (\bibinfo{year}{1983}).

\bibitem[{\citenamefont{Frankfurt and Strikman}(1985)}]{Frankfurt:1985cv}
\bibinfo{author}{\bibfnamefont{L.}~\bibnamefont{Frankfurt}} \bibnamefont{and}
  \bibinfo{author}{\bibfnamefont{M.}~\bibnamefont{Strikman}},
  \bibinfo{journal}{Nucl. Phys.} \textbf{\bibinfo{volume}{B250}},
  \bibinfo{pages}{143} (\bibinfo{year}{1985}).

\bibitem[{\citenamefont{Frankfurt and Strikman}(1988)}]{Frankfurt:1988nt}
\bibinfo{author}{\bibfnamefont{L.}~\bibnamefont{Frankfurt}} \bibnamefont{and}
  \bibinfo{author}{\bibfnamefont{M.}~\bibnamefont{Strikman}},
  \bibinfo{journal}{Phys. Rep.} \textbf{\bibinfo{volume}{160}},
  \bibinfo{pages}{235} (\bibinfo{year}{1988}).

\bibitem[{\citenamefont{Gross and Liuti}(1992)}]{Gross:1991pi}
\bibinfo{author}{\bibfnamefont{F.}~\bibnamefont{Gross}} \bibnamefont{and}
  \bibinfo{author}{\bibfnamefont{S.}~\bibnamefont{Liuti}},
  \bibinfo{journal}{Phys. Rev. C} \textbf{\bibinfo{volume}{45}},
  \bibinfo{pages}{1374} (\bibinfo{year}{1992}).

\bibitem[{\citenamefont{Frank et~al.}(1996)\citenamefont{Frank, Jennings, and
  Miller}}]{Frank:1995pv}
\bibinfo{author}{\bibfnamefont{M.~R.} \bibnamefont{Frank}},
  \bibinfo{author}{\bibfnamefont{B.~K.} \bibnamefont{Jennings}},
  \bibnamefont{and} \bibinfo{author}{\bibfnamefont{G.~A.}
  \bibnamefont{Miller}}, \bibinfo{journal}{Phys. Rev. C}
  \textbf{\bibinfo{volume}{54}}, \bibinfo{pages}{920} (\bibinfo{year}{1996}).

\bibitem[{\citenamefont{Boffi et~al.}(1993)\citenamefont{Boffi, Giusti, and
  Pacati}}]{Boffi:1993gs}
\bibinfo{author}{\bibfnamefont{S.}~\bibnamefont{Boffi}},
  \bibinfo{author}{\bibfnamefont{C.}~\bibnamefont{Giusti}}, \bibnamefont{and}
  \bibinfo{author}{\bibfnamefont{F.}~\bibnamefont{Pacati}},
  \bibinfo{journal}{Phys. Rep.} \textbf{\bibinfo{volume}{226}},
  \bibinfo{pages}{1} (\bibinfo{year}{1993}).

\bibitem[{\citenamefont{Alekhin}(2003)}]{Alekhin:2002fv}
\bibinfo{author}{\bibfnamefont{S.}~\bibnamefont{Alekhin}},
  \bibinfo{journal}{Phys. Rev. D} \textbf{\bibinfo{volume}{68}},
  \bibinfo{pages}{014002} (\bibinfo{year}{2003}).

\bibitem[{\citenamefont{Bodek et~al.}(1979)}]{PhysRevD.20.1471}
\bibinfo{author}{\bibfnamefont{A.}~\bibnamefont{Bodek}} \bibnamefont{et~al.},
  \bibinfo{journal}{Phys. Rev. D} \textbf{\bibinfo{volume}{20}},
  \bibinfo{pages}{1471} (\bibinfo{year}{1979}).

\bibitem[{\citenamefont{Christy and Bosted}(2010)}]{Christy:2007ve}
\bibinfo{author}{\bibfnamefont{M.~E.} \bibnamefont{Christy}} \bibnamefont{and}
  \bibinfo{author}{\bibfnamefont{P.~E.} \bibnamefont{Bosted}},
  \bibinfo{journal}{Phys. Rev. C} \textbf{\bibinfo{volume}{81}},
  \bibinfo{pages}{055213} (\bibinfo{year}{2010}).

\bibitem[{\citenamefont{Lacombe et~al.}(1980)}]{Lacombe:1980dr}
\bibinfo{author}{\bibfnamefont{M.}~\bibnamefont{Lacombe}} \bibnamefont{et~al.},
  \bibinfo{journal}{Phys. Rev. C} \textbf{\bibinfo{volume}{21}},
  \bibinfo{pages}{861} (\bibinfo{year}{1980}).

\bibitem[{\citenamefont{Machleidt}(2001)}]{Machleidt:2000ge}
\bibinfo{author}{\bibfnamefont{R.}~\bibnamefont{Machleidt}},
  \bibinfo{journal}{Phys. Rev. C} \textbf{\bibinfo{volume}{63}},
  \bibinfo{pages}{024001} (\bibinfo{year}{2001}).

\bibitem[{\citenamefont{Gross and Stadler}(2008)}]{Gross:2008ps}
\bibinfo{author}{\bibfnamefont{F.}~\bibnamefont{Gross}} \bibnamefont{and}
  \bibinfo{author}{\bibfnamefont{A.}~\bibnamefont{Stadler}},
  \bibinfo{journal}{Phys. Rev. C} \textbf{\bibinfo{volume}{78}},
  \bibinfo{pages}{014005} (\bibinfo{year}{2008}).

\bibitem[{\citenamefont{Frankfurt et~al.}(1993)\citenamefont{Frankfurt,
  Strikman, Day, and Sargsian}}]{Frankfurt:1993sp}
\bibinfo{author}{\bibfnamefont{L.}~\bibnamefont{Frankfurt}},
  \bibinfo{author}{\bibfnamefont{M.}~\bibnamefont{Strikman}},
  \bibinfo{author}{\bibfnamefont{D.}~\bibnamefont{Day}}, \bibnamefont{and}
  \bibinfo{author}{\bibfnamefont{M.}~\bibnamefont{Sargsian}},
  \bibinfo{journal}{Phys. Rev. C} \textbf{\bibinfo{volume}{48}},
  \bibinfo{pages}{2451} (\bibinfo{year}{1993}).

\bibitem[{\citenamefont{Bueltmann et~al.}(2006)\citenamefont{Bueltmann,
  Christy, Fenker, Griffioen, Keppel, Kuhn, and Melnitchouk}}]{BONUS12}
\bibinfo{author}{\bibfnamefont{S.}~\bibnamefont{Bueltmann}},
  \bibinfo{author}{\bibfnamefont{M.~E.} \bibnamefont{Christy}},
  \bibinfo{author}{\bibfnamefont{H.}~\bibnamefont{Fenker}},
  \bibinfo{author}{\bibfnamefont{K.}~\bibnamefont{Griffioen}},
  \bibinfo{author}{\bibfnamefont{C.~E.} \bibnamefont{Keppel}},
  \bibinfo{author}{\bibfnamefont{S.}~\bibnamefont{Kuhn}}, \bibnamefont{and}
  \bibinfo{author}{\bibfnamefont{W.}~\bibnamefont{Melnitchouk}},
  \bibinfo{journal}{Jefferson Lab Experiment E12-10-102 [BONUS12]}
  (\bibinfo{year}{2006}).

\end{thebibliography}

\end{document}